\documentstyle[12pt,dina4,amsfont12]{article} 

\input psfig.sty 
\begin{document} 

\newcommand{\pl}[2]{\frac{\partial#1}{\partial#2}} 
\newcommand{\Ta}{{\cal A}} 
\newcommand{\Tb}{{\cal B}} 
\newcommand{\Te}{{\cal E}} 
\newcommand{\bu}{{\bf u} } 
\newcommand{\p}{\partial} 
\newcommand{\og}{\omega} 
\newcommand{\Og}{\Omega} 
\newcommand{\fl}[2]{\frac{#1}{#2}} 
\newcommand{\dt}{\delta} 
\newcommand{\tm}{\times} 
\newcommand{\sm}{\setminus} 
\newcommand{\nn}{\nonumber} 
\newcommand{\ap}{\alpha} 
\newcommand{\bt}{\beta} 
\newcommand{\ld}{\lambda} 
\newcommand{\Gm}{\Gamma} 
\newcommand{\gm}{\gamma} 
\newcommand{\vp}{\varphi} 
\newcommand{\tht}{\theta} 
\newcommand{\ift}{\infty} 
\newcommand{\vep}{\varepsilon} 
\newcommand{\ep}{\epsilon} 
\newcommand{\kp}{\kappa} 
\newcommand{\Dt}{\Delta} 
\newcommand{\Sg}{\Sigma} 
\newcommand{\fa}{\forall} 
\newcommand{\sg}{\sigma} 
\newcommand{\ept}{\emptyset} 
\newcommand{\btd}{\nabla} 
\newcommand{\btu}{\Delta} 
\newcommand{\tg}{\triangle} 
\newcommand{\Th}{{\cal T}_h} 
\newcommand{\ged}{\qquad \Box} 
\newcommand{\bgv}{\Bigg\vert} 
\renewcommand{\theequation}{\arabic{section}.\arabic{equation}} 
\newcommand{\be}{\begin{equation}} 
\newcommand{\ee}{\end{equation}} 
\newcommand{\ba}{\begin{array}} 
\newcommand{\ea}{\end{array}} 
\newcommand{\bea}{\begin{eqnarray}} 
\newcommand{\eea}{\end{eqnarray}} 
\newcommand{\beas}{\begin{eqnarray*}} 
\newcommand{\eeas}{\end{eqnarray*}} 
\newcommand{\dpm}{\displaystyle} 
\newtheorem{theorem}{Theorem}[section] 
\newtheorem{lemma}{Lemma}[section] 
\newtheorem{remark}{Remark}[section] 
\newcommand{\Gmu}{\Gm_{_U}} 
\newcommand{\Gml}{\Gm_{_L}} 
\newcommand{\Gme}{\Gm_e} 
\newcommand{\Gmi}{\Gm_i} 
\newcommand{\lN}{{_N}} 
\newcommand{\tld}[1]{\~{#1}} 
\newcommand{\td}[1]{\tilde{#1}} 
\newcommand{\um}{\mu} 
\newcommand{\dts}{\stackrel{\cdot}{*}} 

\newcommand{\bPsi}{{\bf \Psi} } 
\newcommand{\bPhi}{{\bf \Phi} } 
\newcommand{\bx}{{\bf x} } 
\newcommand{\bU}{{\bf U} } 
\newcommand{\bV}{{\bf V} } 
\newcommand{\bz}{{\bf 0} } 
\newcommand{\bA}{{\bf A} } 
\newcommand{\bC}{{\bf C} } 
\newcommand{\bmu}{{\mathcal U} } 
 
\title{Ground states and dynamics of multi-component 
 Bose-Einstein condensates} 
\author{ \ 
{\it Weizhu Bao} 
\thanks{Email address:  bao@cz3.nus.edu.sg.}\\ 
Department of Computational Science\\ 
National University of Singapore, Singapore 117543\\ 
} 
 
\date{} 
\maketitle 
 
\begin{abstract} 
We study numerically the time-independent vector Gross-Pitaevskii
equations (VGPEs) for ground states and time-dependent VGPEs
with (or without) an external driven field for dynamics describing a 
multi-component Bose-Einstein condensate (BEC) at zero or very 
low temperature. In preparation for the numerics, we scale the 3d VGPEs,
approximately reduce it to lower dimensions,
present a normalized gradient flow (NGF) to compute 
ground states of multi-component BEC, 
prove energy diminishing of the NGF which 
provides a mathematical justification, discretize 
it by the backward Euler finite difference (BEFD) which 
is monotone in linear and nonlinear cases
and preserves energy diminishing property in linear case.
Then we use a time-splitting sine-spectral method (TSSP) to discretize the 
time-dependent VGPEs with an external driven field for computing 
dynamics of multi-component BEC. The merit of the
TSSP for VGPEs is that it is explicit, unconditionally stable,
time reversible and time transverse invariant if the VGPEs is, 
`good' resolution in the semiclassical regime, 
of spectral order accuracy in space and second order accuracy in time, 
and conserves the total particle number in the discretized level. 
Extensive numerical examples in 3d for ground states and dynamics 
of multi-component BEC are presented to demonstrate the power of the 
numerical methods and to discuss the 
physics of multi-component Bose-Einstein condensates.
\end{abstract} 
 
\bigskip

 
\bigskip


\section{Introduction}\label{si} 
\setcounter{equation}{0} 
 
  Since its realization in dilute atomic gases \cite{Anderson,Bradley}, 
Bose-Einstein 
condensation (BEC) of alkali atoms and hydrogen has
been produced and studied extensively in the laboratory \cite{Hall}, 
and has afforded an intriguing glimpse into the
macroscopic quantum world.  In view of potential applications, such 
as the generation of bright beams of coherent matter waves (atom laser),
a central goal has been the formation of condensate with the
number of atoms as large as possible. It is thus of particular 
interest to study a scenario where this goal is achieved by 
uniting two (or more) independently grown condensates to form
one large single condensate. The first experiment involving the
uniting of  multiple-component BEC was performed with 
atoms evaporately cooled in the $|F=2, m_f=2\rangle$ and $|1, -1\rangle$
spin states of $^{87}$Rb \cite{Myatt}. Physically speaking, 
two independently formed condensates are characterized by a random
relative phase of their macroscopic wave functions.
A ``fusing'' of two condensates thus amounts to locking the relative
phase in a dissipative process. 
Currently, there are two typical ways 
to lock the relative phase: (i). An external driven field \cite{Myatt};
(ii). An internal atomic Josephson junction \cite{Jaksch}.    
In fact, recent experimental advances in exploration 
of systems of uniting two or more condensates, 
e.g. in a magnetic trap in rubidium 
\cite{Myatt} and subsequently in an optical trap in Sodium \cite{Stamper}, 
have spurred great excitement in the atomic physics 
community and renewed interest in studying the ground states
and dynamics of multi-component BEC \cite{Hall,Chen,Jaksch,Courteille}.

   Theoretical treatment of such systems  began 
in the context of superfluid helium 
mixtures \cite{Khalatnikov} and spinpolarized hydrogen \cite{Siggia}, and
has now been extended to BEC in the alkalis \cite{Ho,Esry,Law,Pu}.
Theoretical predications of properties of uniting two or more condensates,
e.g. density profile, dynamics of interacting BEC condensates 
\cite{Goldstein},
motional damping \cite{Jaksch} and formation of vortices 
\cite{Isoshima,Kita,Leanhard},
can now be compared with experimental data \cite{Hall,Anglin}.
Needless to say that this dramatic progress on the experimental front 
has stimulated a wave of activity on both the theoretical
and the numerical front. In fact, the properties of uniting two 
or more BEC states at temperatures $T$ much smaller than
the critical condensation temperature $T_c$ \cite{LL} are usually modeled by 
the vector Gross-Pitaevskii equations (VGPSs) for the macroscopic vector wave 
function \cite{Pit,LL} either with an external driven field \cite{Hall} or
with an internal  atomic Josephson junction \cite{Jaksch}. 
Note that equations very similar to the VGPEs also appear
in nonlinear optics 
where indices of refraction, which depend on the light 
intensity, leads to nonlinear terms 
like those encountered in VGPEs.

  There have been extensive numerical studies of the time-independent 
Gross-Pitaevskii equation (GPE) for ground states and
time-dependent GPE for dynamics of single-component BEC.
For computing ground states of BEC, Bao and Du \cite{BD} presented
a normalized gradient flow (NGF), proved energy diminishing and 
discretized it by a backward Euler finite difference (BEFD) method;
Bao and Tang \cite{Bao} proposed a method which can be used 
to compute the ground and excited states via directly minimizing 
the energy functional; Edwards and Burnett \cite{Edwards} introduced 
a Runge-Kutta type method; other methods include
an explicit imaginary-time algorithm 
used in \cite{Adu} and \cite{Tosi}; a directly inversion in the iterated 
subspace (DIIS) used in  Schneider et al. \cite{Feder}
and a simple analytical type method proposes by Dodd \cite{Dodd}.
For numerical solutions of time-dependent GPE for 
finding dynamics of BEC, Bao et al. \cite{Bao3} presented 
a time-splitting spectral (TSSP) method, Ruprecht et al. \cite{Rup}
used the Crank-Nicolson finite difference method, Cerimele 
et al. \cite{Tosi2,Cerim}
proposed a particle-inspired scheme. Up to now, only a few 
numerical simulations on multi-component BEC \cite{Jaksch,Chen,Ghosh}.

  In this paper, we take the 3d vector Gross-Pitaevskii equations 
(VGPEs) with an external driven field for multi-component BEC, 
make it dimensionless, 
approximately reduce it to a 2d VGPEs and a 1d VGPEs in certain limits,
and discuss the approximate ground state solution of VGPEs 
in (very) weak interaction regime. Then we present a 
normalized gradient flow (NGF) to compute 
ground states of multi-component BEC, 
prove energy diminishing of the NGF which 
provides a mathematical justification, discretize 
it by the backward Euler finite difference (BEFD) which 
is monotone in linear and nonlinear cases
and preserves energy diminishing property in linear case.
At last, we use a time-splitting sine-spectral method (TSSP),
which was studied in Bao et al. \cite{Bao1,BJP} for the nonlinear
Schr\"{o}dinger equation (NLS) in the semiclassical regime
and used for GPE of single-component BEC \cite{Bao3}, damped GPE 
for collapse and explosion of BEC \cite{Bao4} and Zakharov system for 
plasma physics \cite{BS,BSW},  
 to discretize the 
time-dependent VGPEs with an external driven field for computing 
dynamics of multi-component BEC. The merit of the
TSSP for VGPEs is that it is explicit, unconditionally stable,
easy to program, less memory requirement, 
time reversible and time transverse invariant if the VGPEs is, 
`good' resolution in the semiclassical regime, 
of spectral order accuracy in space and second order accuracy in time, 
and conserves the total particle number in the discretized level.
Extensive numerical examples in 3d for ground states and dynamics 
of multi-component BEC are presented to demonstrate the power of the 
numerical methods.
 
  The paper is organized as follows. In section \ref{sgpel} we start out 
with the 3d VGPEs with an external driven field, make it dimensionless,
show how to reduce it to lower dimensions. In section \ref{sgss}
we give approximate ground
state solution in (very) weak interaction regime,
present a normalized gradient flow (NGF) to compute 
ground states of multi-component BEC, 
prove energy diminishing of the NGF, discretize 
it by the backward Euler finite difference (BEFD), as well as apply 
the NGF and its BEFD discretization  
 to a nonlinear two-state model for vortex states dynamics
in BEC.
In section \ref{stssp}, we present the time-splitting 
sine-spectral (TSSP) method for
the VGPEs with an external driven field. 
In section \ref{sne} numerical 
tests of the VGPEs for ground states and dynamics 
of multi-component BEC are presented. In section \ref{ss} a summary is given.
Throughout we adopt the standard $l^2$-norm of vectors, matrices, 
and $\|\cdot\|$ as the standard $L^2$-norm for functions, as well as
the $\dts$ operator which is used in Matlab for two vectors
$\bU=(u_1,\cdots, u_M)^T$ and $\bV=(v_1,\cdots,v_M)^T$ as
$\bU\dts \bV=(u_1v_1,\cdots, u_M v_M)^T$.

\section{The vector Gross-Pitaevskii equations (VGPEs)}\label{sgpel} 
\setcounter{equation}{0} 

  At temperatures $T$ much smaller than the critical temperature 
$T_c$ \cite{LL}, a BEC for $M$ components with an external 
driven field is well described by the macroscopic vector wave function 
 $\bPsi = \bPsi({\bf x},t)=(\psi_1(\bx,t), \cdots, \psi_M(\bx,t))^T$ 
whose evolution is governed by 
a self-consistent, 
mean field vector  Gross-Pitaevskii equations \cite{Gross,Pit}. 
If the harmonic trap potential is considered, the VGPEs become 
\be 
\label{vgpe} 
i\hbar\pl{\bPsi(\bx,t)}{t}=-\fl{\hbar^2}{2m}\btd^2 \bPsi(\bx,t)+ 
\hat{\bV}(\bx)\dts \bPsi(\bx,t) +\hat{\bA}(\bPsi)\dts \bPsi(\bx,t)
+\hbar \hat{f}(t) \hat{B} \bPsi(\bx,t),
\ee 
where $\bx=(x,y,z)^T$ is the spatial coordinate vector,  
$m$ is the atomic mass, $\hbar=1.05\tm 10^{-34} [J\,s]$ 
is the Planck constant, 
$\hat{f}(t)$ is a given real-valued scalar function, $\hat{B}
=(b_{jl})_{j,l=1}^M$ is a
given $M\tm M$ symmetric real matrix, i.e. $b_{jl}=b_{lj}$ 
($j,l=1,\cdots,M$), 
$\hat{\bV}(\bx)=(\hat{V}_1(\bx),\; \cdots,\; \hat{V}_M(\bx))^T$ 
is the harmonic trap potential, 
i.e.
\[\hat{V}_j(\bx)= \fl{m}{2}\left(\og_{x,j}^2 \; (x-\hat{x}_{0,j})^2+ 
\og_{y,j}^2 \; (y-\hat{y}_{0,j})^2
 +\og_{z,j}^2 \; (z-\hat{z}_{0,j})^2\right), \quad j=1,\cdots, M\]
with $(\hat{x}_{0,j},\hat{y}_{0,j},\hat{z}_{0,j})^T$ and 
$\og_{x,j}$, $\og_{y,j}$, $\og_{z,j}$
the center and trap frequencies in $x$, $y$, and 
$z$-direction, respectively,
of the $j$th ($j=1,\cdots, M$) component. For the following
we assume (w.r.o.g.) $\og_{x,1}\le \cdots \og_{x,M}\le \og_{y,1}\cdots
\le \og_{z,M}$.  $\hat{\bA}(\bPsi)=(\hat{A}_1(\bPsi),\; \cdots, \;
\hat{A}_M(\bPsi))^T$ 
models the 
interaction, i.e. 
\beas
\hat{A}_j(\bPsi)&=&u_{j1}\; |\psi_1|^2+\cdots+u_{jM}\; |\psi_M|^2\\
&=&\fl{4\pi\hbar^2a_{j1}}{m}\; |\psi_1|^2+\cdots +
\fl{4\pi\hbar^2a_{jM}}{m}\; |\psi_M|^2,
\qquad j=1,\cdots,M,
\eeas
%
%
with $u_{jl} =\fl{4\pi\hbar^2 a_{jl}}{m}$ and 
$a_{jl}=a_{lj}$ the $s$-wave scattering length between
the $j$th and $l$th component (positive for repulsive 
interaction and negative for attractive interaction, $j,l=1,\cdots,M$).
It is necessary to ensure that the vector wave function is
properly normalized. Specifically, we require
\be
\label{normi}
\int_{{\Bbb R}^3} |\psi_j(\bx,0)|^2\; d\bx =N_j^0>0, \qquad 
j=1,\cdots, M,
\ee
where $N_j^0$  is the number of 
particles of the $j$th ($j=1\cdots, M$) component at time $t=0$.

\subsection{Dimensionless VGPEs}

In order to scale the VGPEs (\ref{vgpe}),  we introduce 
\be \label{dml} 
\tilde{t}=\og_{x,1} t, \quad \tilde{\bx}=\fl{\bx}{a_0}, \quad 
\tilde{\psi_j}(\tilde \bx,\tilde t)=\fl{a_0^{3/2}}{\sqrt{N_j^0}}
\psi_j(\bx,t), \ j=1,\cdots,M, \quad a_0=\sqrt{\fl{\hbar}{m\og_{x,1}}},
\ee 
where 
$a_0$ is the length of the harmonic oscillator ground state.
In fact, here we choose $1/\og_{x,1}$ and $a_0$ as the dimensionless
time and length units, respectively. Plugging 
(\ref{dml}) into (\ref{vgpe}), multiplying  by $\fl{1}{m \og_{x,1}^2 
(N_j^0 a_0)^{1/2}}$ to the $j$th ($j=1\cdots, M$) 
equation,  and then removing all \~{}, we 
obtain the following dimensionless VGPEs
in 3d with an external driven field 
\be 
\label{gpe2} 
i\;\pl{\bPsi(\bx,t)}{t}=-\fl{1}{2}\btd^2 \bPsi(\bx,t)+ 
\bV(\bx)\dts \bPsi(\bx,t) +\bA(\bPsi)\dts \bPsi(\bx,t)
+f(t) B \bPsi(\bx,t), 
\ee 
where $f(t)=\hat{f}\left(t/\og_{x,1}\right)/\og_{x,1}$, 
and 
\beas
&&\bV(\bx)=(V_1(\bx),\; \cdots,\; V_M(\bx))^T,\\
 &&V_j(\bx)=\fl{1}{2}\left(\gm_{x,j}^2\; 
(x-x_{0,j})^2+\gm_{y,j}^2\; (y-y_{0,j})^2
+ \gm_{z,j}^2 \;(z-z_{0,j})^2\right), \\
&&\gm_{x,j} = \fl{\og_{x,j}}{\og_{x,1}}, \qquad
\gm_{y,j} = \fl{\og_{y,j}}{\og_{x,1}}, \qquad
\gm_{z,j} = \fl{\og_{z,j}}{\og_{x,1}},\\
&&x_{0,j}= \fl{\hat{x}_{0,j}}{a_0}, \qquad 
y_{0,j}= \fl{\hat{y}_{0,j}}{a_0}, \qquad 
z_{0,j}= \fl{\hat{z}_{0,j}}{a_0}, \qquad j=1,\cdots, M,\\
&&\bA(\bPsi)=\left(A_1(\bPsi), \; \cdots,\; A_M(\bPsi)\right)^T,\\ 
&&A_j(\bPsi)=\bt_{j1}|\psi_1|^2+\cdots+\bt_{jM}|\psi_M|^2,
 \qquad j=1,\cdots,M,\\
&&\bt_{jl}= \fl{ u_{jl} N_l^0}{ a_0^3 \hbar \og_{x,1}}
=\fl{4\pi \hbar^2 a_{jl} N_l^0}{m a_0^3 \hbar \og_{x,1}}
=\fl{4\pi a_{jl} N_l^0}{a_0}, 
\quad j,l=1,\cdots, M,\\
&&B=G_0^{-1}\  \hat{B}\ G_0, \qquad \hbox{with}\qquad 
G_0={\rm diag}\left(\sqrt{N_1^0},\; \cdots,\; \sqrt{N_M^0}\right).
\eeas

  The VGPEs (\ref{gpe2}) conserves the {\bf normalization of the vector 
wave function} or the total number of particles  
\be
\label{norm}
N(G_0\bPsi) = \int_{{\Bbb R}^3} \|G_0\bPsi(\bx,t)\|_{l^2}^2\; 
d\bx =\sum_{j=1}^M\int_{{\Bbb R}^3} N_j^0|\psi_j(\bx,t)|^2\;d\bx
=N^0, 
\quad t\ge0.
\ee
When there is no external driven field, i.e. $f\equiv 0$ in (\ref{gpe2}),
the VGPEs (\ref{gpe2}) is time reversible, time transverse invariant,
and conserves the {\bf normalization of the  
wave function for each component} or the number of particles 
of each component
\be
\label{normc}
N_j(\psi_j) = \int_{{\Bbb R}^3} |\psi_j(\bx,t)|^2\; d\bx =1, \qquad t\ge0,
\qquad j=1,\cdots, M
\ee
and the {\bf energy}
\be
\label{energy}
E_\bt(\bPsi)=\sum_{j=1}^M\ \fl{N_j^0}{N^0}\ E_{\bt,j}(\bPsi),
\ee
with
\beas
&&N^0=N_1^0+\cdots+N_M^0, \qquad \bt=\max_{1\le j,l\le M}\ |\bt_{jl}|,\\
&&E_{\bt,j}(\bPsi)=\int_{{\Bbb R}^3} \left[\fl{1}{2}|\btd\psi_j|^2 +
V_j(\bx) |\psi_j|^2+\fl{1}{2}\sum_{l=1}^M\bt_{jl}|\psi_j|^2 |\psi_l|^2
\right]\; d\bx, \quad
j=1,\cdots, M.
\eeas

  There are two extreme regimes: one is when $\bt\ll1$ 
($\Longleftrightarrow$ $|\bt_{jl}|\ll1$ for all
$j,l=1,\cdots,M$), then the system (\ref{gpe2}) describes a weakly 
interacting condensation. The other one
is when $\bt\gg1$, then (\ref{gpe2}) 
corresponds to a strongly interacting condensation or
to the semiclassical regime or the Thomas-Fermi regime.

\subsection{Reduction to lower dimensions} 

  In the following two cases, the 3d VGPEs (\ref{gpe2}) without 
external driven field, i.e. $f\equiv0$,
can approximately be reduced to 2d or even 1d. 
In the case (disk-shaped condensation)
\be
\label{dsc}
\og_{x,j}\approx \og_{y,j}\approx \og_{x,1}, 
\ \og_{z,j}\gg \og_{x,1} \quad \Longleftrightarrow \quad 
\gm_{x,j}\approx \gm_{y,j}\approx 1, \ \gm_{z,j}\gg 1, 
\quad j=1,\cdots,M,
\ee
the 3d VGPEs (\ref{gpe2}) can be reduced to 2d VGPEs 
with $\bx=(x,y)^T$ by assuming that the time evolution does not cause 
excitations along the $z$-axis since they
have large energy of approximately $\hbar \og_{z,j}$ compared
to excitations along the $x$ and $y$-axis with
energies of about $\hbar\og_{x,1}$. Thus we may assume that
the condensate wave function along the $z$-axis is 
always well described by the ground state wave function 
and set 
\be
\label{redw}
\bPsi =\bPsi_2(x,y,t) \dts \bPsi_3(z) 
\quad \hbox{with} \quad \bPsi_3(z) =\left(\int_{{\Bbb R}^2} 
 \bPhi_g(x,y,z) \dts \bPhi_g^*(x,y,z) \;dxdy\right)^{1/2},
\ee
where $\bPsi_2(x,y,t)=\left(\psi_{2,1}(x,y,t),\, \cdots,\, 
\psi_{2,M}(x,y,t)\right)^T$,
$\bPsi_3(z)=\left(\psi_{3,1}(z), \, \cdots, \, \psi_{3,M}(z)\right)^T$ and
$\bPhi_g(x,y,z)=\left(\phi_{g,1}(x,y,z),\, \cdots, \,
\phi_{g,M}(x,y,z)\right)^T$ 
(see detail in (\ref{mine})) is the ground state solution 
of the 3d VGPEs (\ref{gpe2}) by setting $f\equiv0$ and $g^*$ denotes the 
conjugate of a function $g$. Plugging (\ref{redw}) into (\ref{gpe2}),
then $\dts$ both sides by $\bPhi_3^*(z)$, 
integrating with respect to $z$ over $(-\ift,\ift)$, we get
\be 
\label{gpe2d} 
i\;\pl{\bPsi_2(\bx,t)}{t}=-\fl{1}{2}\btd^2 \bPsi_2(\bx,t)+ 
\left(\bV(\bx)+\bC\right)\dts \bPsi_2(\bx,t) 
+\bA(\bPsi_2)\dts \bPsi_2(\bx,t), 
\ee 
where 
\beas
&&\bV(\bx)=\left(V_1(x,y),\; \cdots,\; V_M(x,y)\right)^T,\\
&&V_j(x,y)= \fl{1}{2}\left(\gm_{x,j}^2\; (x-x_{0,j})^2 
+\gm_{y,j}^2\; (y-y_{0,j})^2\right), 
\qquad j=1,\cdots,M,\\
&&\bC=\left(c_1,\; \cdots,\; c_M\right)^T,\\
&&c_j = \fl{\gm_{z,j}^2}{2} \int_{-\ift}^\ift (z-z_{0,j})^2 
|\psi_{3,j}|^2 \; dz +
\fl{1}{2}\int_{-\ift}^\ift \left|\fl{d\psi_{3,j}(z)}{dz}\right|^2\;dz,
\qquad j=1,\cdots, M,\\
&&\bA(\bPsi)=\left(A_1(\bPsi),\; \cdots,\; A_M(\bPsi)\right)^T,\\
&&A_j(\bPsi)=\sum_{l=1}^M \left(\bt_{jl}\int_{-\ift}^\ift 
|\psi_{3,j}(z)|^2 |\psi_{3,l}(z)|^2\;dz\right)
|\psi_{2,l}|^2, \qquad j=1,\cdots,M.
\eeas
Since this VGPEs is time-transverse invariant, we can replace
$\bPsi_2\to \bPsi_2 \dts e^{-i\bC t/2}$ which drops the 
constant vector $\bC$ in the trap potential and obtain 
the 2d VGPEs with $\bPsi = \bPsi_2$ and $\bx=(x,y)^T$
\be 
\label{gpe2dd} 
i\;\pl{\bPsi(\bx,t)}{t}=-\fl{1}{2}\btd^2 \bPsi(\bx,t)+ 
\bV(\bx)\dts \bPsi(\bx,t) +\bA(\bPsi)\dts \bPsi(\bx,t).
\ee 
The observables are not affected by this.

Similarly in the case (cigar-shaped condensation)
\be
\label{dscc}
\og_{x,j}\approx \og_{x,1},\ \og_{y,j}\gg \og_{x,1}, 
\  \og_{z,j}\gg \og_{x,1} \ \Longleftrightarrow \ 
\gm_{x,j}\approx 1, \ \gm_{y,j}\gg 1, \ \gm_{z,j}\gg 1, 
\ 1\le j\le M,
\ee
the 3d VGPEs (\ref{gpe2}) can be reduced to 1d VGPEs 
with $\bx=x$. Similarly to the 2d case, we derive the 1d VGPEs
\be 
\label{gpe1dd} 
i\;\pl{\bPsi(x,t)}{t}=-\fl{1}{2}\btd^2 \bPsi(x,t)+ 
\bV(x)\dts \bPsi(x,t) +\bA(\bPsi)\dts \bPsi(x,t),
\ee 
where
\beas
&&\bV(x)=\left(V_1(x),\; \cdots,\; V_M(x)\right)^T,\\
&&V_j(x)= \fl{1}{2}\gm_{x,j}^2\; (x-x_{0,j})^2, 
\qquad j=1,\cdots,M,\\
&&\bA(\bPsi)=\left(A_1(\bPsi),\; \cdots,\; A_M(\bPsi)\right)^T,\\
&&A_j(\bPsi)=\sum_{l=1}^M\left(\bt_{jl}\int_{{\Bbb R}^2} 
|\psi_{23,j}(y,z)|^2\;
|\psi_{23,l}(y,z)|^2\;dz\right) |\psi_l|^2,\\
&&\psi_{23,j}(y,z)=\left(\int_{-\ift}^\ift |\phi_{g,j}(x,y,z)|^2\; 
dx\right)^{1/2}, \qquad j=1,\cdots, M.
\eeas

In fact, the 3d VGPEs (\ref{gpe2}), 2d VGPEs (\ref{gpe2dd}) 
and 1d VGPEs (\ref{gpe1dd})
with an external driven field can be written in a unified way
\be 
\label{gpeu}
i\;\pl{\bPsi(\bx,t)}{t}=-\fl{1}{2}\btd^2 \bPsi(\bx,t)+ 
\bV_d(\bx)\dts \bPsi(\bx,t) +\bA_d(\bPsi)\dts \bPsi(\bx,t)
+f(t) B \bPsi(\bx,t),  \  \bx \in {\Bbb R}^d,
\ee 
where
\beas
&&\bV_d(\bx)=\left(V_{d,1}(\bx),\; \cdots,\; V_{d,M}(\bx)\right)^T,\\
&&\bA_d(\bPsi)=\left(A_{d,1}(\bPsi),\; \cdots,\; A_{d,M}(\bPsi)\right)^T,\\
&&A_{d,j}(\bPsi)=\bt_{d,j1}\;|\psi_1|^2+\cdots+\bt_{d,jM}\;|\psi_M|^2,
 \qquad j=1,\cdots,M;
\eeas
with 
\beas
&&V_{d,j}=\left\{\ba{ll}
\fl{1}{2}\gm_{x,j}^2\; (x-x_{0,j})^2, &\quad d=1,\\
\fl{1}{2}\left(\gm_{x,j}^2\;(x-x_{0,j})^2+\gm_{y,j}^2\; (y-y_{0,j})^2\right), 
  &\quad d=2,\\
\fl{1}{2}\left(\gm_{x,j}^2\;(x-x_{0,j})^2+\gm_{y,j}^2\; (y-y_{0,j})^2
+\gm_{z,j}^2\; (z-z_{0,j})^2\right), &\quad d=3;\\
\ea\right. \\
&&\bt_{d,jl}=\left\{\ba{ll}
\bt_{jl}\int_{{\Bbb R}^2}|\psi_{23,j}|^2\ |\psi_{23,l}|^2\; dydz, 
   &\qquad\qquad d=1,\\
\bt_{jl}\int_{-\ift}^\ift|\psi_{3,j}|^2\ |\psi_{3,l}|^2\; dz, 
&\qquad\qquad d=2,\\
\bt_{jl}, &\qquad\qquad d=3.\\
\ea\right.
\eeas
 The VGPEs (\ref{gpeu}) conserves the {\bf normalization of the vector 
wave function} or the total number of particles  
\be
\label{normmg}
N(G_0\bPsi) = \int_{{\Bbb R}^d} \|G_0\bPsi(\bx,t)\|_{l^2}^2\; d\bx 
= \sum_{j=1}^M \int_{{\Bbb R}^d} N_j^0|\psi_j(\bx,t)|^2\; d\bx 
=N^0, \qquad t\ge0.
\ee
When there is no external driven field, i.e. $f\equiv0$ in (\ref{gpeu}),
the VGPEs (\ref{gpeu}) is time reversible, time transverse invariant,
and conserves the {\bf normalization of the  
wave function for each component} or the number of particles 
of each component
\be
\label{normcc}
N_j(\psi_j) = \int_{{\Bbb R}^d} |\psi_j(\bx,t)|^2\; d\bx =1, \qquad t\ge0,
\qquad j=1,\cdots, M
\ee
and the {\bf energy}
\be
\label{energyy}
E_\bt(\bPsi)=\sum_{j=1}^M\ \fl{N_j^0}{N^0}\ E_{\bt,j}(\bPsi),
\ee
with
\[
E_{\bt,j}(\bPsi)=\int_{{\Bbb R}^d} \left[\fl{1}{2}|\btd\psi_j|^2 +
V_{d,j}(\bx) |\psi_j|^2+\fl{1}{2}\sum_{l=1}^M\bt_{d, jl}\;|\psi_j|^2\; 
|\psi_l|^2 \right]\; d\bx, \quad
j=1,\cdots, M.
\]

\section{Ground state solution}\label{sgss} 
\setcounter{equation}{0}

To find a stationary solution of (\ref{gpeu}) without external driven
field, i.e. $f\equiv0$, we write
\be
\label{ans}
\bPsi(\bx,t) = e^{-i\; \bmu \; t} \dts \bPhi(\bx), 
\ee
where $\bmu=\left(\mu_1, \cdots, \mu_M\right)^T$ is the 
chemical potential vector of the multi-component condensate and 
$\Phi(\bx)=\left(\phi_1(\bx), \cdots, \phi_M(\bx)\right)^T$ a 
real-valued vector
function independent of time. Inserting (\ref{ans}) into (\ref{gpeu}) 
gives the following equations for $\left(\bmu, \bPhi\right)$:
\be
\label{engp}
\bmu\dts \bPhi(\bx)= -\fl{1}{2}\btu \bPhi(\bx) + \bV_d(\bx)\dts
\bPhi(\bx) + \bA_d(\bPhi)\dts \bPhi(\bx), 
\qquad \bx \in {\Bbb R}^d,
\ee
under the normalization condition
\be
\label{normep}
\int_{{\Bbb R}^d} \left|\phi_j(\bx)\right|^2 \; d\bx =1, 
\qquad j=1,\cdots,M.
\ee
This is a nonlinear eigenvalue problem under the constraint
(\ref{normep}) and any eigenvalue vector $\bmu$ can be computed 
from its corresponding eigenfunction vector $\bPhi$ by
\bea
\label{eigenv}
\mu_j&=&\mu_{\bt,j}(\bPhi)=\int_{{\Bbb R}^d}
\left[\fl{1}{2}|\btd \phi_j(\bx)|^2 +V_{d,j}(\bx)|\phi_j(\bx)|^2 +
\bA_{d,j}(\bPhi) |\phi_j(\bx)|^2\right]\; d\bx \nn\\
&=&\int_{{\Bbb R}^d}
\left[\fl{1}{2}|\btd \phi_j(\bx)|^2 +V_{d,j}(\bx)|\phi_j(\bx)|^2 +
\sum_{l=1}^M \bt_{d,jl}\;|\phi_l(\bx)|^2\; |\phi_j(\bx)|^2\right]\; d\bx \nn\\
&=&E_{\bt,j}(\bPhi) +\fl{1}{2}\int_{{\Bbb R}^d} 
\sum_{l=1}^M \bt_{d,jl}\; |\phi_l(\bx)|^2\; |\phi_j(\bx)|^2\; d\bx,
\qquad j=1,\cdots,M.
\eea
It is easy to see that critical points of the energy functional 
$E_\bt(\bPhi)$ under the constraint (\ref{normep}) are eigenfunctions
of the nonlinear eigenvalue problem 
(\ref{engp}) under the constraint (\ref{normep}) and versus versa.
In fact, (\ref{engp}) can be viewed as the Euler-Lagrange
equations of the energy functional $E_\bt(\bPhi)$ under the
constraint (\ref{normep}).  
The multi-component BEC ground state solution
$\bPhi_g(\bx)$ is found by 
minimizing the energy $E_\bt(\bPhi)$ under the constraint
(\ref{normep}), i.e.

\noindent (V) Find $\left( \bmu_g=(\mu_{g,1},\; \cdots,\; \mu_{g,M})^T,
\ \bPhi_g=(\phi_{g,1},\;\cdots,\; \phi_{g,M})^T\in \bU\right)$ such that 
\bea
\label{mine}
&&E_\bt(\bPhi_g)=\min_{\bPhi\in \bU}\ E_\bt (\bPhi), \\
&&\mu_{g,j}=\mu_{\bt,j}(\bPhi_g) =
E_{\bt,j}(\bPhi_g)+\fl{1}{2}\int_{{\Bbb R}^d} 
 \sum_{l=1}^M\bt_{d, jl}\; |\phi_{g,j}(\bx)|^2\;
 |\phi_{g,l}(\bx)|^2\; d\bx,  \ 1\le j\le M, \qquad\quad  
\eea
where the set $\bU$ is defined as
\[\bU=\left\{ \bPhi \ |\  E_\bt(\bPhi) <\ift, \quad 
\int_{{\Bbb R}^d} |\phi_j(\bx)|^2 \; d\bx=1, \ 1\le j\le M\right\}. \]

In non-rotating multi-component BEC, the minimization problem 
(\ref{mine}) has a unique real-valued nonnegative 
ground state solution $\bPhi_g(\bx)>\bz$ for $\bx\in {\Bbb R}^d$
\cite{Lieb}.  When $M=1$, i.e. single-component BEC, 
the minimizer of (\ref{mine}) was computed by either
a normalized gradient flow \cite{BD}, or directly minimizing 
the energy functional \cite{Bao}, or the imaginary time method
\cite{Adu,Tosi}, etc. Here we extend the normalized gradient flow method
for computing ground state solution from single-component 
BEC to multi-component BEC.

\subsection{Normalized gradient flow and energy diminishing}

Consider the following continuous normalized gradient flow
\bea
\label{ngf1}
&&\bPhi_t= \fl{1}{2}\btu \bPhi -\bV_d(\bx)\dts \bPhi - 
\bA_d(\bPhi)\dts \bPhi+\bmu_\bPhi(t) \dts \bPhi, 
\quad \bx \in {\Bbb R}^d, \ t\ge0,\\
\label{ngf2}
&&\bPhi(\bx,0)=\bPhi_0(\bx)=\left(\phi_{0,1}(\bx),\; \cdots,\;
\phi_{0,M}(\bx)\right)^T,  \qquad \bx \in {\Bbb R}^d;
\eea
where $\bmu_\bPhi(t)=\left(\mu_{\bPhi,1}(t),\; \cdots, \;
\mu_{\bPhi,M}(t)\right)^T$ with 
\bea
\label{ngf3}
\bmu_{\bPhi,j}(t)&=&\fl{1}{\|\phi_j(\cdot,t)\|^2} 
\int_{{\Bbb R}^d} \left[\fl{1}{2} |\btd \phi_j(\bx,t)|^2 
+V_{d,j}(\bx)\; |\phi_j(\bx,t)|^2 \right. \nn\\
&&\qquad \qquad \left. +  \sum_{l=1}^M\bt_{d, jl}\; |\phi_{l}(\bx,t)|^2\; 
|\phi_j(\bx,t)|^2\right]\;d\bx, \quad j=1,\cdots, M.\qquad 
\eea
In fact, the right hand side of (\ref{ngf1}) is the same as
(\ref{engp}) if we view $\bmu_\bPhi(t)$ as a Lagrange multiplier for the
constraint (\ref{normep}). Furthermore, as observed in \cite{BD}
for single-component BEC, the solution of (\ref{ngf1}) also satisfies the 
following theorem:

  \begin{theorem}\label{edhh}
Suppose $\bV_d(\bx)\ge\bz$ for all $\bx \in {\Bbb R}^d$, 
$\bt_{jl}\ge0$  ($j,l=1,\cdots,M$) and 
$\|\phi_{0,j}\|=1$ ($j=1,\cdots,M$). Then the normalized gradient flow 
(\ref{ngf1})-(\ref{ngf2}) is normalization conservation and
energy diminishing, i.e.
\bea
\label{ncphi}
&&\|\phi_j(\cdot,t)\|^2=\int_{{\Bbb R}^d} \phi_j^2(\bx,t)\; d\bx =
\|\phi_{0,j}\|^2=1, \qquad t\ge0, \quad j=1,\cdots,M,\\
\label{edcngf}
&&
\fl{d}{dt}E_\bt(\bPhi)=-\sum_{j=1}^M \fl{2N_j^0}{N^0}
 \left\|\p_t \phi_j(\cdot,t)\right\|^2
=-\sum_{j=1}^M \fl{2N_j^0}{N^0} \int_{{\Bbb R}^d} 
\left|\p_t \phi_j(\bx,t)\right|^2\;d\bx \le 0\;, 
\ t\ge0,\qquad    \qquad 
\eea
which in turn implies
$$
E_\bt(\bPhi(\cdot, t_1))\ge E_\bt(\bPhi(\cdot,t_2)), 
\qquad 0\le t_1\le t_2<\ift.
$$
\end{theorem}

\bigskip

\noindent Proof: Multiplying the $j$th ($j=1,\cdots,M$) equation
in (\ref{ngf1}) by $\phi_j$, integrating over ${\Bbb R}^d$, integration
by parts and notice (\ref{ngf3}), we obtain
\bea
\label{jcon}
\lefteqn{\fl{1}{2}\fl{d}{dt}\int_{{\Bbb R}^d} 
\phi_j^2(\bx,t)\;d\bx =
\int_{{\Bbb R}^d}  \phi_j\; \p_t \phi_j\; d\bx  }\nn\\[2mm]
& = & \int_{{\Bbb R}^d} \left[\fl{1}{2}\btu \phi_j - V_{d,j}(\bx) \phi_j 
-A_{d,j}(\bPhi)\phi_j
+\um_{\bPhi,j}(t)\phi_j\right]\phi_j\;d\bx
 \nn\\[2mm]
&=&-\int_{{\Bbb R}^d} \left[\fl{1}{2}|\btd \phi_j(\bx,t)|^2+V_{d,j}(\bx)
\phi_j^2(\bx,t)+A_{d,j}(\bPhi)\phi_j^2\right]d\bx 
+\um_{\bPhi,j}(t) \|\phi_j(\cdot,t)\|^2 \nn\\
&=&0, \qquad t\ge0, \qquad j=1,\cdots,M.
\eea
This implies the normalization conservation (\ref{ncphi}).

Next, direct calculation shows
\bea
\label{decngf}
\lefteqn{\fl{d}{dt}E_\bt(\bPhi)=\sum_{j=1}^M \fl{N_j^0}{N^0}\ 
\fl{d}{dt}E_{\bt,j}(\bPhi)}\nn\\[2mm]
&=&\sum_{j=1}^M \fl{N_j^0}{N^0} \int_{{\Bbb R}^d} \left[
\btd\phi_j \cdot \btd (\p_t \phi_j) 
+2V_{d,j}(\bx) \phi_j \p_t\phi_j +
\sum_{l=1}^M \bt_{jl} \left(|\phi_l|^2\phi_j\p_t \phi_j  + |\phi_j|^2
\phi_l \p_t \phi_l\right) \right]d\bx \nn \\ 
&=&\sum_{j=1}^M \fl{N_j^0}{N^0} \int_{{\Bbb R}^d} \left[
\btd\phi_j \cdot \btd (\p_t \phi_j) 
+2V_{d,j}(\bx) \phi_j \p_t\phi_j +
\sum_{l=1}^M \bt_{jl} |\phi_l|^2\phi_j\p_t \phi_j \right]d\bx \nn\\
&&+\sum_{j=1}^M \sum_{l=1}^M \fl{N_j^0}{N^0}\int_{{\Bbb R}^d}
\bt_{jl}|\phi_j|^2 \phi_l \p_t \phi_l \; d\bx \nn\\
&=&\sum_{j=1}^M \fl{N_j^0}{N^0} \int_{{\Bbb R}^d} \left[
\btd\phi_j \cdot \btd (\p_t \phi_j) 
+2V_{d,j}(\bx) \phi_j \p_t\phi_j +
\sum_{l=1}^M \bt_{jl}  |\phi_l|^2\phi_j\p_t \phi_j\right]d\bx \nn\\
&&+\sum_{l=1}^M \fl{N_l^0}{N^0}\sum_{j=1}^M \int_{{\Bbb R}^d}
\bt_{lj}|\phi_j|^2\phi_l \p_t \phi_l \; d\bx \nn\\
&=&\sum_{j=1}^M \fl{2N_j^0}{N^0} \int_{{\Bbb R}^d} \left[
\btd\phi_j \cdot \btd (\p_t \phi_j) 
+2V_{d,j}(\bx) \phi_j \p_t\phi_j +
\sum_{l=1}^M \bt_{jl}|\phi_l|^2 \phi_j\p_t \phi_j \right]d\bx \nn\\
&=&2 \int_{{\Bbb R}^d}\left[-\fl{1}{2}\btu \phi_j +V_{d,j}(\bx) \phi_j 
+A_{d,j}(\bPhi)\phi_j \right]\p_t \phi_j
\; d\bx \nn   \\
&=&\sum_{j=1}^M \fl{2N_j^0}{N^0}\int_{{\Bbb R}^d} \left[-\p_t\phi_j(\bx,t) 
+\mu_{\bPhi,j}(t) \phi_j(\bx,t)\right] \p_t\phi_j\;d\bx \nn   \\ 
&=&-\sum_{j=1}^M \fl{2N_j^0}{N^0}\|\p_t\phi_j(\cdot, t) \|^2 
+\mu_{\bPhi,j}(t)\; \fl{d}{dt} \int_{{\Bbb R}^d} |\phi_j(\bx,t)|^2\;d\bx
\nn \\
&=&-\sum_{j=1}^M \fl{2N_j^0}{N^0} \|\p_t\phi_j(\cdot, t) \|^2 \;,
 \qquad t\ge0,
\eea
since $\mu_{\bPhi,j}(t)$ ($j=1,\cdots,M$) are always real and
$$
\fl{d}{dt} \int_{{\Bbb R}^d}  |\phi_j(\bx,t)|^2\;d\bx =0, 
\qquad j=1,\cdots,M
$$
due to the normalization conservation. Thus, we easily get
$$
E_\bt(\bPhi(\cdot, t_1))\ge E_\bt(\bPhi(\cdot,t_2)), 
\qquad 0\le t_1\le t_2<\ift
$$
for the solution of (\ref{ngf1}). \hfill $\Box$

\bigskip

Using argument similar to that in \cite{Simon}, we may also
get as $t\to\ift$, $\bPhi$ approaches to a steady state solution
which is a critical point of the energy.  In non-rotating multi-component
BEC, it has a unique real valued nonnegative 
ground state solution $\bPhi_g(\bx)\ge0$ 
for all $\bx\in{\Bbb R}^d$  \cite{Lieb}.
We choose the 
initial data $\bPhi_0(\bx)\ge0$ for $\bx\in{\Bbb R}^d$, e.g. 
the approximate ground state solution (\ref{engcl}) in weakly interacting
multi-component BEC. 
Under this kind of initial data, 
the ground state solution $\bPhi_g$ and its corresponding chemical
potential $\bmu_g$ can be obtained from 
the steady state solution of the normalized gradient flow 
(\ref{ngf1})-(\ref{ngf2}), i.e. 
\bea
\label{gstlime}
&&\bPhi_g(\bx)=\lim_{t\to\ift} \bPhi(\bx,t),\quad \bx\in{\Bbb R}^d, \\ 
\label{gstlimv}
&&\mu_{g,j} =\mu_{\bt,j}(\bPhi_g)=E_\bt(\bPhi_g)
+\fl{1}{2}\int_{{\Bbb R}^d } \sum_{l=1}^M\bt_{d, jl}|\phi_{g,l}(\bx)|^2
  |\phi_{g,j}(\bx)|^2\; d\bx,
\ j=1,\cdots, M. \qquad \quad 
\eea

\subsection{Projection}

When one wants to evolve the normalized gradient flow (\ref{ngf1}),
(\ref{ngf2}) numerically, 
it is natural to consider the following 
projection (or splitting) scheme which was widely used in physical 
literatures for computing the ground state solution 
of single-component BEC \cite{BD} by constructing a time sequence
$0=t_0<t_1<t_2<\cdots <t_n<\cdots$ with $t_n=n\;k$ and $k>0$ time step:
\bea
\label{ngf1p}
&&\bPhi_t=
\fl{1}{2}\btu \bPhi -\bV_d(\bx)\dts \bPhi - 
\bA_d(\bPhi)\dts \bPhi, 
\quad \bx \in {\Bbb R}^d, \ t_n\le t<t_{n+1},\ n\ge0, \qquad \quad \\
\label{ngf2p}
&&\phi_j(\bx,t_{n+1})\stackrel{\triangle}{=} 
\phi_j(\bx,t_{n+1}^+)=\fl{\phi_j(\bx,t_{n+1}^-)}
{\|\phi_j(\cdot,t_{n+1}^-)\|}, 
\qquad \bx\in {\Bbb R}^d, \quad n\ge 0,\\
\label{ngf3p}
&&\bPhi(\bx,0)=\bPhi_0(\bx),  \qquad \bx \in {\Bbb R}^d;
\eea
where $\bPhi(\bx,t_n^\pm)=\left(\phi_1(\bx,t_n^\pm), \cdots, 
\phi_M(\bx,t_n^\pm)\right)^T=\lim_{t\to t_n^\pm} 
\bPhi(\bx,t)$ and $\|\phi_{0,j}\|=1$ ($j=1,\cdots,M$).
In fact, the  gradient flow with projection (\ref{ngf1p}),
(\ref{ngf2p}) can be viewed as applying the steepest decent 
method to the minimization problem (\ref{mine}) by ignoring 
the constraint $\bPhi\in \bU$ and then projecting back to
the set $\bU$. The  gradient flow (\ref{ngf1p})
can also be viewed as applying an imaginary time 
(i.e. $t\to -i t$) in (\ref{gpeu}). 
The normalized step (\ref{ngf2p}) is equivalent to solving
the following ODE system {\sl exactly}
\bea
\label{Ode1}
&&\bPhi_t(\bx,t) = \bmu_\bPhi(t,k)\dts \bPhi(\bx,t), 
\qquad \bx\in{\Bbb R}^d,
\quad t_n \le t<t_{n+1}, \quad n\ge0,\\
\label{Ode2}
&&\bPhi(\bx,t_n^+)= \bPhi(\bx,t_{n+1}^-), \qquad   \bx\in{\Bbb R}^d;
\eea
where $\bmu_\bPhi(t,k)=\left(\mu_{\bPhi,1}(t,k),\cdots,
\mu_{\bPhi,M}(t,k)\right)^T$ with 
\be
\label{sgtk}
\mu_{\bPhi,j}(t,k)\equiv
\um_{\bPhi,j}(t_{n+1},k) = -\fl{1}{2\; k}
\ln \|\phi_j(\cdot,t_{n+1}^-)\|^2, 
\quad t_n\le t\le t_{n+1}, \quad j=1,\cdots,M.
\ee
Thus the  gradient flow with projection
can be viewed as a first-order
splitting method for the following continuous gradient flow with 
discontinuous coefficients: 
\bea
\label{nngf1}
&&\bPhi_t=
\fl{1}{2}\btu \bPhi -\bV_d(\bx)\dts \bPhi - 
\bA_d(\bPhi)\dts \bPhi+ \bmu_\bPhi(t,k)\dts \bPhi, 
\quad \bx \in {\Bbb R}^d, \ n\ge0,\qquad \\
\label{nngf2}
&&\bPhi(\bx,0)=\bPhi_0(\bx),  \qquad \bx \in {\Bbb R}^d;
\eea
Let $k\to 0$, note (\ref{sgtk}) and (\ref{ngf1p}), we see that 
\bea
\um_{\bPhi,j}(t)&=&\lim_{k\to0^+}\um_{\bPhi,j}(t,k)
=\fl{1}{\|\phi_j(\cdot,t)\|^2}\int_{{\Bbb R}^d}
\left[\fl{1}{2}|\btd \phi_j(\bx,t)|^2+V_{d,j}(\bx)|\phi_j(\bx,t)|^2
 \right. \nn\\
&&\qquad \quad \left. + \sum_{l=1}^M\bt_{d, jl}\;|\phi_{l}(\bx,t)|^2\;
 |\phi_j(\bx,t)|^2 \right]d\bx, \qquad j=1,\cdots, M,
\eea
which implies that  the problem of (\ref{nngf1}), (\ref{nngf2}) collapses to
(\ref{ngf1p}), (\ref{ngf2p}) as $k\to 0$. Furthermore,
using the Theorem 2.1 
in \cite{BD}, we get immediately

\begin{theorem}\label{edh}
Suppose $\bV_d(\bx)\ge\bz$ for all $\bx \in {\Bbb R}^d$ and 
$\|\phi_{0,j}\|=1$ ($j=1,\cdots, M$).  For $\bt_{jl}=0$ 
($j,l=1,\cdots,M$),  the 
gradient flow with projection (\ref{ngf1p})-(\ref{ngf3p}) 
is energy diminishing under any time step $k$ and
initial data $\bPhi_0$, i.e.
\be
\label{dphi}
E_0(\bPhi(\cdot,t_{n+1}))\le E_0(\bPhi(\cdot, t_n))\le \cdots
\le E_0(\bPhi(\cdot,0))=E_0(\bPhi_0), \qquad n=0,1,2,\cdots.
\ee
\end{theorem}

\subsection{Backward Euler finite difference (BEFD) discretization}

In this subsection, we present a fully discretization of the 
 gradient flow with projection (\ref{ngf1p}), (\ref{ngf2p}) by 
the backward Euler finite difference (BEFD) which was proposed
in \cite{BD} for discretizing a normalized gradient flow
for single-component BEC.

 For simplicity of 
notation we shall introduce the method for the case of one spatial 
dimension $(d=1)$ with 
homogeneous Dirichlet boundary conditions.
 Generalizations to $d>1$ are straightforward 
for tensor product grids and the results remain valid without 
modifications. For $d=1$, the problem becomes 
\begin{eqnarray} \label{sdge1d} 
&&\bPhi_t = \fl{1}{2}\bPhi_{xx} - \bV_1(x)\dts \bPhi - 
\bA_1(\bPhi)\dts \bPhi, \quad 
a< x < b,\ t_n\le t<t_{n+1}, \ n\ge0,\qquad \\
\label{sdge1d2}
&&\phi_j(x,t_{n+1})\stackrel{\triangle}{=}
\phi_j(x,t_{n+1}^+)=\fl{\phi_j(x,t_{n+1}^-)}{\|\phi_j(\cdot,t_{n+1}^-)\|}, 
\quad a\le x\le b, \ n\ge 0, \ 
j=1,\cdots,M,  \qquad \quad \\
\label{sdge1d3}
&&\bPhi(x,0)=\bPhi_0(x), \qquad a\le x \le b,\\
\label{sdge1d4}
&&\bPhi(a,t)=\bPhi(b,t)=\bz, \qquad t\ge0;
\eea
with 
\[\|\phi_{0,j}\|^2=\int_a^b \phi_{0,j}^2(x)\; dx=1, \qquad j=1,\cdots,M.\]

  We choose the spatial mesh size $h=\btu x>0$ with $h=(b-a)/N$ and $N$ 
an even positive integer, and define grid points by 
\[ x_j:= a+j\; h, \qquad j=0,1,\cdots,N.\]
Let $\bPhi_j^n=((\phi_1)_j^n, \; \cdots,\; (\phi_M)_j^n)^T$ 
be the numerical approximation of $\bPhi(x_j,t_n)=(\phi_1(x_j,t_n),\;$, 
$\cdots$, $\phi_M(x_j,t_n))^T$.
Here we use the backward Euler for time discretization and second-order
centered finite difference for spatial derivatives for the
 gradient flow (\ref{ngf1p}). The detail scheme is:
\bea
&&\fl{\bPhi_j^*-\bPhi_j^n}{k}=\fl{1}{2h^2}\left[\bPhi_{j+1}^*
-2\bPhi_j^*+\bPhi_{j-1}^*\right]-\bV_1(x_j)\dts \bPhi_j^*
-\bA_1(\bPhi^n_j) \dts\bPhi_j^*,  \nn\\
&&\qquad \quad \qquad \quad \qquad \quad j=1,\cdots, N-1,\nn\\
&&\bPhi_0^*=\bPhi_N^*=0,\nn\\
\label{befd3}
&&(\phi_l)_j^{n+1}=\fl{(\phi_l)_j^*}{\sqrt{h\sum_{s=1}^{N-1} 
\left((\phi_l)_s^*\right)^2}}, 
\quad j=0,\cdots, N, \ l=1,\cdots,M, \ 
\ n=0,1,\cdots,\qquad \quad \\
&&\bPhi_j^0= \bPhi_0(x_j), \qquad j=0,1,\cdots, N. \nn
\eea

  It is easy to see that the discretizetion BEFD (\ref{befd3}) 
is monotone for any time step $k>0$ when $\bV_1(\bx)\ge \bz$ and
$\bt_{jl}\ge0$ ($j,l=1,\cdots,M$). Furthermore, similar to the proof
of Theorem 3.1 in \cite{BD}, we can prove the BEFD normalized flow
(\ref{befd3}) is energy diminishing  
for any time step $k>0$ when $\bV_1(\bx)\ge \bz$ and
$\bt_{jl}=0$ ($j,l=1,\cdots,M$).

\begin{remark} 
 Extension of the BEFD discretization (\ref{sdge1d}) for mutli-component
BEC can be done as those in the Appendix in \cite{BD}
for single-component BEC  in the cases when 
$\bV_d(\bx)$ in 2d with radial symmetry or
in 3d with spherical symmetry or cylindrical symmetry, as 
well as in 2d or 3d for central vortex states.
\end{remark}

\subsection{Approximate ground state solution}

For  a weakly interacting condensate, i.e. $\bt\ll1$ ($\Longleftrightarrow$
$|\bt_{jl}|\ll1$, $j,l=1,\cdots,M$), we drop the nonlinear terms (i.e. the
last term on the right hand side of (\ref{engp})) and
find the linear vector Schr\"{o}dinger equations with
the harmonic oscillator potentials
\be
\label{engpl}
\bmu\dts \bPhi(\bx)= -\fl{1}{2}\btu \bPhi(\bx) + \bV_d(\bx)\dts
\bPhi(\bx), 
\qquad \bx=(x_1,\cdots,x_d)^T \in {\Bbb R}^d,
\ee
under the normalization condition (\ref{normep}). The 
ground state solution of (\ref{engpl}) is \cite{Levine}
\bea
\label{engvl}
&&\mu_{g,j}^w = \fl{\gm_{x_1,j}+\cdots+\gm_{x_d,j}}{2}, \qquad
 j=1,\cdots, M, \\
\label{engcl}
&&\phi_{g,j}^w(\bx)=\fl{\left(\gm_{x_1,j}\cdots \gm_{x_d,j}\right)^{1/4}}
{\pi^{d/4}}\ e^{-(\gm_{x_1,j} (x_1- (x_1)_{0,j})^2+\cdots+
\gm_{x_d,j} (x_d- (x_d)_{0,j})^2)/2}.
\eea
It can be viewed as an approximate ground state solution of 
(\ref{engp}) in the case of a weakly interacting multi-component BEC. 
This approximate ground state can be used as initial data
in the normalized gradient flow (\ref{ngf1}), or (\ref{ngf1p})
and (\ref{ngf2p}), or (\ref{nngf1}) for  
computing the ground state solution of multi-component
BEC when $\bt_{jl}\ne0$.

\subsection{Application to a two-state model}

  The normalized gradient flow method and its BEFD discretization 
for multi-component BEC can be applied to compute coupled 
basis wavefunctions with lowest energy 
 of the nonlinear two-state model used in \cite{CD1,CD2}
for studying vortex dynamics in single-component BEC with (or without) an 
external rotation. For the continence of the reader, here
we briefly review the derivation of the nonlinear two-state 
model from the Gross-Pitaevskii equation (GPE). Consider the dimensionless 
GPE for BEC in 2d with radial symmetry \cite{BD,Bao,Bao3}:
\be
\label{gpe2dt}
i\; \psi_t(r,\tht, t) = -\fl{1}{2}\left[\fl{1}{r}\pl{}{r}
\left(r\pl{\psi}{r}\right)
+\fl{\p^2 \psi}{\p\tht^2}\right] +\fl{r^2}{2} \psi +\bt |\psi|^2\psi, 
\ee
under the normalization condition
\[\int_0^\ift \int_0^{2\pi} |\psi(r,\tht,t)|^2\; r\; drd\tht=1,
\]
where $(r,\tht)$ is the polar coordinate, $\psi(r,\tht,t)$ is the 
macroscopic wave function for the condensate, $\bt$ is a parameter 
models the interaction. In order to represent the condensate mean-field 
wavefunction $\psi$ by the superposition of a symmetric component $\phi_s$
and a vortex component $\phi_v e^{i\tht}$, we 
take the ansatz
\be
\label{tsm}
\psi(r,\tht,t) = a_s \phi_s(r;n_v) e^{-i\mu_s t}
+ a_v \phi_v(r;n_v) e^{i\tht} e^{-i\mu_v t},
\ee
where $a_s$ and $a_v$ are the complex amplitudes of the symmetric 
and vortex components, respectively. The vortex fraction is 
$0\le n_v=|a_v|^2\le 1$ and the symmetric fraction is $n_s=|a_s|^2 =1-n_v$.
The $\phi_s$ and $\phi_v$ are real nonnegative functions, and are
normalized to unity, i.e.
\be
\label{contsm}
2\pi \int_0^\ift |\phi_s(r;n_v)|^2\; r\; dr =1,
\qquad 2\pi \int_0^\ift |\phi_v(r;n_v)|^2\; r\; dr =1. 
\ee
Plugging (\ref{tsm}) into (\ref{gpe2d}), multiplying both sides
by $1$ and $e^{-i\tht}$, respectively, and then integrating
over ${\Bbb R}^2$, see detail in \cite{CD1}, we get the following 
nonlinear two-state model
\bea
\label{tsmse}
&&\mu_s \phi_s = -\fl{1}{2r}\fl{d}{dr} \left(r\fl{d\phi_s}{dr}\right)
 +\fl{r^2}{2}\phi_s + \bt\left(n_s \phi_s^2 + 2F n_v \phi_v^2\right) \phi_s,\\
\label{tsmve}
&&\mu_v \phi_v = -\fl{1}{2r}\fl{d}{dr} \left(r\fl{d\phi_v}{dr}\right)
 +\left(\fl{r^2}{2}+\fl{1}{2r^2}\right)\phi_v + \bt\left(2Fn_s \phi_s^2 
+ n_v \phi_v^2\right) \phi_v, \\
&&\left.\fl{d\phi_s(r;n_v)}{dt}\right|_{r=0}=0, 
\ \phi_v(0;n_v)=0, \
\lim_{r\to\ift}\phi_s(r;n_v)=\lim_{r\to\ift}\phi_v(r;n_v)=0;\qquad
\eea
where the factor $F=1$, but can be adjusted in some cases. 
In order to study vortex dynamics 
in BEC through the two-state model (\ref{tsmse}),  (\ref{tsmve}) 
\cite{CD1,CD2},
one needs to find the coupled basis wavefunctions
$\phi_s(r;n_v)$ and $\phi_v(r;n_v)$ for any given $0\le n_v\le 1$ 
by minimizing the energy $E(\phi_s,\phi_v)$ defined as
\bea
\label{egtsm}
&&E(\phi_s,\phi_v)= n_s\; E_s(\phi_s,\phi_v) +n_v\; E_v(\phi_s,\phi_v),\\
&&E_s(\phi_s,\phi_v)=2\pi \int_0^\ift \fl{r}{2}\left[\left|
\fl{d\phi_s(r;n_v)}{dr}\right|^2 + r^2\phi_s^2(r;n_v)\right.\nn\\
&&\qquad \qquad \qquad\qquad \left.+\bt\left( n_s \phi_s^2(r;n_v) 
+ 2F n_v \phi_v^2(r;n_v)\right)
\phi_s^2(r;n_v) \right]\;dr, \nn\\
&&E_v(\phi_s,\phi_v)=2\pi \int_0^\ift \fl{r}{2}\left[\left|
\fl{d\phi_v(r;n_v)}{dr}\right|^2 + \left(r^2+\fl{1}{r^2}\right)\phi_v^2(r;n_v)
\right. \nn\\
&&\qquad\qquad  \qquad\qquad \left.+\bt\left( 2Fn_s \phi_s^2(r;n_v) 
+ n_v \phi_v^2(r;n_v)\right)
\phi_v^2(r;n_v) \right]\;dr;\nn\\
\eea
under the constraint (\ref{contsm}). The continuous 
normalized gradient flow for computing the above minimizer is:
\bea
\label{tsngf1}
&&\fl{\p \phi_s(r,t;n_v)}{\p t} 
= \fl{1}{2r}\fl{d}{dr} \left(r\fl{d\phi_s}{dr}\right)
 -\fl{r^2}{2}\phi_s - \bt\left(n_s \phi_s^2 + 2F n_v \phi_v^2\right) \phi_s
+\mu_s(t)\phi_s ,\\
\label{tsngf2}
&&\fl{\p \phi_v(r,t;n_v)}{\p t} = 
\fl{1}{2r}\fl{d}{dr} \left(r\fl{d\phi_v}{dr}\right)
 -\left(\fl{r^2}{2}+\fl{1}{2r^2}\right)\phi_v
 - \bt\left(2Fn_s \phi_s^2 
+ n_v \phi_v^2\right) \phi_v +\mu_v(t)\phi_v, \qquad \qquad \\
\label{tsngf3}
&&\left.\fl{\p\phi_s(r,t;n_v)}{\p t}\right|_{r=0}=0, 
\ \phi_v(0,t;n_v)=0, \ 
\lim_{r\to\ift}\phi_s(r,t;n_v)=\lim_{r\to\ift}\phi_v(r,t;n_v)=0,\qquad \\
\label{tsngf4}
&&\phi_s(r,0;n_v)=\phi_{s,0}(r)\ge0, 
\qquad \phi_v(r,0;n_v)=\phi_{v,0}(r)\ge0, \qquad 0\le r<\ift;
\eea
with 
\[2\pi \int_0^\ift |\phi_{s,0}(r)|^2\; r\; dr =1,
\qquad 2\pi \int_0^\ift |\phi_{v,0}(r)|^2 \;r\; dr =1 
\]
and 
\beas
&&\mu_{s}(t) = \fl{1}{\int_0^\ift r\;|\phi_s(r,t;n_v)|^2\;dr}
 \int_0^\ift \fl{r}{2}\left[\left|
\fl{d\phi_{s}(r,t;n_v)}{dr}\right|^2 + r^2\phi_{s}^2(r,t;n_v)\right.\nn\\ 
&&\qquad \qquad\qquad \left.+2\bt\left( n_s \phi_{s}^2(r,t;n_v) 
+ 2F n_v \phi_{v}^2(r,t;n_v)\right)
\phi_{s}^2(r,t;n_v) \right]\;dr,\nn\\
&&\mu_{v}(t)= \fl{1}{\int_0^\ift r\;|\phi_v(r,t;n_v)|^2\;dr}
\int_0^\ift \fl{r}{2}\left[\left|
\fl{d\phi_{v}(r,t;n_v)}{dr}\right|^2 + 
\left(r^2+\fl{1}{r^2}\right)\phi_{v}^2(r,t;n_v) \right.\nn\\
&&\qquad \qquad\qquad \left.+2\bt\left( 2Fn_s \phi_{s}^2(r,t;n_v) 
+ n_v \phi_{v}^2(r,t;n_v)\right)
\phi_{v}^2(r,t;n_v) \right]\;dr.
\eeas
If we choose the initial data $\phi_{s,0}(r)\ge0$ and $\phi_{v,0}(r)\ge0$
for $0\le r<\ift$, e.g. $\phi_{s,0}(r)=\fl{1}{\pi^{1/2}} e^{-r^2/2}$
and $ \phi_{v,0}(r)=\fl{r}{\pi^{1/2}} e^{-r^2/2}$, 
then the minimizer $\phi_{s,g}(r;n_v)$ and $\phi_{v,g}(r;n_v)$
can be obtained from the steady state solution 
of the normalized gradient flow (\ref{tsngf1})-(\ref{tsngf4}), i.e.
\beas
&&\phi_{s,g}(r;n_v)=\lim_{t\to \ift} \phi_{s}(r,t;n_v),
\quad \phi_{v,g}(r;n_v)=\lim_{t\to \ift} \phi_{v}(r,t;n_v),
\quad 0\le r<\ift; \qquad \quad \\
&&\mu_{s,g} = 2\pi \int_0^\ift \fl{r}{2}\left[\left|
\fl{d\phi_{s,g}(r;n_v)}{dr}\right|^2 + r^2\phi_{s,g}^2(r;n_v)\right.\nn\\ 
&&\qquad \qquad\qquad \left.+2\bt\left( n_s \phi_{s,g}^2(r;n_v) 
+ 2F n_v \phi_{v,g}^2(r;n_v)\right)
\phi_{s,g}^2(r;n_v) \right]\;dr,\nn\\
&&\mu_{v,g}=2\pi \int_0^\ift \fl{r}{2}\left[\left|
\fl{d\phi_{v,g}(r;n_v)}{dr}\right|^2 + 
\left(r^2+\fl{1}{r^2}\right)\phi_{v,g}^2(r;n_v) \right.\nn\\
&&\qquad \qquad\qquad \left.+2\bt\left( 2Fn_s \phi_{s,g}^2(r;n_v) 
+ n_v \phi_{v,g}^2(r;n_v)\right)
\phi_{v,g}^2(r;n_v) \right]\;dr.
\eeas

  Choose $R>0$ sufficiently large, denote the mesh size $h_r=(R-0)/N$,
grid points $r_j = j\; h_r$, $j=0,1,\cdots, N$, and
$r_{j-\fl{1}{2}}=\left(j-\fl{1}{2}\right)h_r$, $j=0,1,\cdots, N+1$.
Then the BEFD discretization of the normalized gradient flow 
(\ref{tsngf1})-(\ref{tsngf4}) reads:
{\small \bea
\label{tsmbe}
&&\fl{(\phi_s)_{j-\fl{1}{2}}^*-(\phi_s)_{j-\fl{1}{2}}^n}{k}=
\fl{1}{2\;h_r^2\;r_{j-\fl{1}{2}}}\left[r_j\;
(\phi_s)_{j+\fl{1}{2}}^*
 -\left(r_j+r_{j-1}\right)(\phi_s)_{j-\fl{1}{2}}^*
+r_{j-1}(\phi_s)_{j-\fl{3}{2}}^*\right]
 -\fl{r_{j-\fl{1}{2}}^2}{2}(\phi_s)_{j-\fl{1}{2}}^*\nn\\
&&\qquad -\bt\left[n_s ((\phi_s)_{j-\fl{1}{2}}^n)^2 + 2F n_v 
\left(\fl{(\phi_v)_{j-1}^n+(\phi_v)_{j}^n}{2}\right)^2\right] 
(\phi_s)_{j-\fl{1}{2}}^*,  \quad j=1,\cdots, N,\nn\\
&&\fl{(\phi_v)_{j}^*-(\phi_v)_j^n}{k}=
\fl{1}{2\;h_r^2\;r_{j}}\left[r_{j+\fl{1}{2}}\;(\phi_v)_{j+1}^*
 -\left(r_{j+\fl{1}{2}}+r_{j-\fl{1}{2}}\right)(\phi_v)_{j}^*
+r_{j-\fl{1}{2}}\;(\phi_v)_{j-1}^*\right]
-\left(\fl{r_{j}^2}{2}+\fl{1}{2r_j^2}\right)\;(\phi_v)_{j}^*\nn\\
&&\qquad -\bt\left[2F n_s \left(\fl{(\phi_s)_{j-\fl{1}{2}}^n
+(\phi_s)_{j+\fl{1}{2}}^n}{2}\right)^2 + n_v 
((\phi_v)_{j}^n)^2\right] (\phi_v)_{j}^*,  \quad j=1,\cdots, N-1,\nn\\
&&(\phi_s)_{-\fl{1}{2}}^*=(\phi_s)_{\fl{1}{2}}^*,\qquad
(\phi_s)_{N+\fl{1}{2}}^*=0, \qquad (\phi_v)_0^*=(\phi_v)_N^*=0,\nn\\
&&(\phi_s)_{j-\fl{1}{2}}^{n+1}=\fl{(\phi_s)_{j-\fl{1}{2}}^*}
{\sqrt{h_r 2\pi \sum_{l=1}^N r_{j-\fl{1}{2}}\;
\left((\phi_s)_{l-\fl{1}{2}}^*\right)^2}}, \qquad j=0,\cdots, N+1, 
\qquad n=0,1,\cdots,\\
\label{tsmce}
&&(\phi_v)_{j}^{n+1}=\fl{(\phi_v)_{j}^*}
{\sqrt{2\pi h_r \sum_{l=1}^{N-1} r_{j}\;
\left((\phi_s)_{j}^*\right)^2} }, \qquad j=0,\cdots, N, 
\qquad n=0,1,\cdots,\\
&&(\phi_s)_{j-\fl{1}{2}}^0= \phi_{s,0}(r_{j-\fl{1}{2}}), 
\qquad j=1,\cdots, N+1,
\qquad (\phi_s)_{-\fl{1}{2}}^0=(\phi_s)_{\fl{1}{2}}^0, \nn\\
&&(\phi_v)_{j}^0= \phi_{v,0}(r_j), \qquad j=0,1,\cdots, N. \nn
\eea }

\begin{remark}
The normalized gradient flow and its BEFD discretization
for the two-state model in 2d with radial symmetry
 can be easily extended to the two-state model in \cite{CD2}
in 3d with cylindrical symmetry.
\end{remark}

\section{TSSP method for dynamics}\label{stssp} 
\setcounter{equation}{0}  

In this section we present a time-splitting sine-spectral (TSSP)
method for the VGPEs (\ref{gpeu}) with (or without) an external driven field
for dynamics of multi-component BEC. 
For simplicity of notation  we shall introduce the method 
in one space dimension $(d=1)$. Generalizations to $d>1$ are 
straightforward for tensor product grids and the results remain 
valid without modifications. For $d=1$, the equations (\ref{gpe2}) 
with homogeneous Dirichlet  boundary conditions become 
\bea \label{sdge1db} 
&&i\;\pl{\bPsi(x,t)}{t}=-\fl{1}{2}\bPsi_{xx}(x,t)+ 
\bV_1(x)\dts \bPsi(x,t) +\bA_1(\bPsi)\dts \bPsi(x,t)
+f(t) B \bPsi(x,t),\nn\\
&&\qquad \qquad \qquad \qquad \quad a<x<b, \ t\ge0,\\
\label{sdgi1d} 
&&\bPsi(a,t)=\bPsi(b,t)=\bz, \qquad t\ge 0,\qquad \\
\label{sdgb1d} 
&&\bPsi(x,t=0)=\bPsi_0(x)=(\psi_{0,1}(x), \; \cdots,\; 
\psi_{0,M}(x))^T, \qquad  a\le x\le b,  
\eea 
with 
\[\|\psi_{0,j}\|= \int_a^b|\psi_{0,j}(x)|^2\;dx =1, \qquad j=1,\cdots,M.\]

We choose the spatial mesh size $h=\btu x>0$ with $h=(b-a)/N$ for $N$ an 
even positive integer, the time step $k=\btu t>0$ 
and let  the grid points and the time step be 
\[ x_j:=a+j\;h, \qquad t_n := n\; k, \qquad j=0,1,\cdots, N, \qquad 
n=0,1,2,\cdots   \] 
Let $\bPsi^n_j=((\psi_1)_j^n,\; \cdots,\; (\psi_M)_j^n)^T$ 
be the approximation of $\bPsi(x_j,t_n)=(\psi_1(x_j,t_n),$ $\;
\cdots,\;$ $\psi_M(x_j,t_n))^T$. 

From time $t=t_n$ to 
$t=t_{n+1}$,  the VGPEs (\ref{sdge1db}) is solved in three splitting 
steps. One solves first 
\be \label{fstep} 
i\;\pl{\bPsi(x,t)}{t}=-\fl{1}{2}\; \bPsi_{xx}(x,t)
 \ee 
for the time step of length $k$, followed by solving 
\be \label{sstep} 
i\;\pl{\bPsi(x,t)}{t}= \bV_1(x)\dts \bPsi(x,t) +\bA_1(\bPsi(x,t))
\dts \bPsi(x,t), 
\ee 
for the same time step, and then by solving 
\be \label{sstep3}
i\;\pl{\bPsi(x,t)}{t}=f(t) \; B \; \bPsi(x,t),
\ee 
Equations (\ref{fstep}) will be discretized in space by the sine-spectral 
method and integrated in time {\it exactly}. For 
$t\in[t_n,t_{n+1}]$, the ODE system  (\ref{sstep}) leaves 
$|\psi_j(x,t)|$ ($j=1,\cdots,M$) invariant in $t$
and  therefore  becomes 
\be 
\label{sstepp} 
i\;\pl{\bPsi(x,t)}{t}= \bV_1(x)\dts \bPsi(x,t) +\bA_1(\bPsi(x,t_n))
\dts \bPsi(x,t), 
 \ee 
and thus can be integrated {\it exactly}. The solution of (\ref{sstepp}) is
\be
\label{step3}
\bPsi(x,t)=e^{-i (\bV_1(x) +\bA_1(\bPsi(x,t_n)))(t-t_n)}\dts \bPsi(x,t_n),
\qquad t_n\le t\le t_{n+1}.
\ee
For 
$t\in[t_n,t_{n+1}]$, the ODE system  (\ref{sstep3})
can be solve {\bf exactly} too. In fact, sine $\hat{B}$ is real 
and symmetric, there exist an orthonormal real   matrix $P$ with
$P^T=P^{-1}$ and a diagonal 
matrix $D={\rm diag}(d_1, \cdots, d_M)$  such that
\[\hat{B}= P\ D \ P^T.\]
Thus the matrix $B$ in (\ref{gpe2}) can be diagonalizable too, i.e.
\be
\label{diag}
B=G_0^{-1}\;P\ D \ P^T\;G_0. 
\ee
Let 
\be
\label{transf}
\bPhi(x,t)= P^T\;G_0\ \bPsi(x,t),
\ee
Plugging (\ref{transf}) into (\ref{sstep3}), noting (\ref{diag}), we get
\be
\label{ephi}
i\;\pl{\bPhi(x,t)}{t}=f(t)\ D\ \bPhi(x,t).
\ee
The solution of (\ref{ephi}) is
\be
\label{sephi}
\bPhi(x,t) = e^{-i\;D\; \int_{t_n}^t f(s)\; ds} \; \bPhi(x,t_n), 
\qquad t_n \le t\le t_{n+1}.
\ee
Substituting (\ref{sephi}) into (\ref{transf}), we obtain the 
solution of the ODE system (\ref{sstep3}) 
\bea
\label{sestep3}
\bPsi(x,t)&=&G_0^{-1}\; P\; e^{-i\;D \int_{t_n}^t f(s)\; ds} \; \bPhi(x,t_n)
\nn \\
&=&G_0^{-1}\; P\; e^{-i\;D \int_{t_n}^t f(s)\; ds} \; 
P^T \;G_0\; \bPsi(x,t_n), \qquad t_n \le t\le t_{n+1}.
\eea
From time $t=t_n$ to $t=t_{n+1}$, we combine the 
splitting steps via the standard second-order splitting: 
\bea 
\label{tsspm}
&&\bPsi_j^{(1)}=\fl{2}{N}\sum_{l=1}^{N-1} 
  e^{-i\;k\;\mu_l^2/4}\;\widehat{(\bPsi^n)}_l\; 
\sin\left(\fl{i\;j\;l\;\pi}{N}\right), \nn\\ 
&&\bPsi_j^{(2)}=e^{-i(\bV_1(x_j) +\bA_1(\bPsi_j^{(1)}))k/2} 
 \dts \bPsi_j^{(1)}, \nn\\
&&\bPsi_j^{(3)} = G_0^{-1}\; P\; e^{-iD \int_{t_n}^{t_{n+1}} 
f(s)\; ds} \; P^T \; G_0\; \bPsi_j^{(2)}, \nn\\
&&\bPsi_j^{(4)}=e^{-i(\bV_1(x_j) +\bA_1(\bPsi_j^{(3)}))k/2} 
 \dts \bPsi_j^{(3)}, \nn\\
&&\bPsi_j^{n+1}=\fl{2}{N}\sum_{l=1}^{N-1} 
  e^{-i\;k\;\mu_l^2/4}\;\widehat{(\bPsi^{(4)})}_l\; 
\sin\left(\fl{i\;j\;l\;\pi}{N}\right),\qquad 1\le j\le N-1;
\eea 
where $\widehat{\bPsi}_l=(\widehat{(\psi_1)}_l,\; \cdots,\;
\widehat{(\psi_M)}_l)^T$ ($l=1,\cdots,N-1$), the sine
coefficients of $\bPsi$ with $\bPsi_0=\bPsi_N=\bz$, are defined as 
\be 
\label{Fouv1} 
\mu_l=\fl{l\pi}{b-a},\quad \widehat{\bPsi}_l=\sum_{j=1}^{N-1} 
 \bPsi_j\;\sin\left(\fl{i\;j\;l\;\pi}{N}\right),  
\quad l=1,\cdots,N-1. 
\ee 
The overall time discretization error comes solely from the splitting, 
which is second order in $k$, and
the spatial discretization is of spectral (i.e. `infinite') order 
of accuracy. It is time reversible and
 time-transverse invariant if the VGPEs (\ref{gpe2}) is, i.e. $f\equiv0$. 
Furthermore, for the stability of the TSSP (\ref{tsspm}), 
we have the following lemma, which shows that
the total number of particles in the mult-component BEC is conserved for any 
given real-valued external driven field $f$, and  the number
of particles of each component is conserved when 
there is no external driven field, i.e. $f\equiv0$.

\begin{lemma} The time-splitting sine-spectral (TSSP) method
(\ref{tsspm}) is unconditionally stable and conserves the total 
number of particles in the mult-component BEC. In fact,
for every mesh size $h>0$ and time step $k>0$,
\bea
\label{consd}
\|G_0 \bPsi^{n}\|_{l^2}&:=&\sqrt{\sum_{l=1}^M N_l^0\; \|(\psi_l)^n\|_{l^2}^2}
= \sqrt{\sum_{l=1}^M N_l^0\; h\sum_{j=1}^{N-1} \left|(\psi_l)_j^n\right|^2}
 \nn\\
&=&\|G_0 \bPsi_{0}\|_{l^2} = \sqrt{\sum_{l=1}^M N_l^0}=\sqrt{N^0},
\qquad n=1,2,\cdots\;.
\eea
Furthermore, when $f\equiv0$ in (\ref{gpe2}), 
i.e. without external driven field,
 we have
\be
\label{consdd}
\|(\psi_l)^n\|_{l^2}:=\sqrt{h\sum_{j=1}^{N-1} \left|(\psi_l)_j^n\right|^2}
=\|(\psi_l)^0\|_{l^2}=1, \qquad n=1,\cdots, \quad j=1,\cdots,M.
\ee
\end{lemma}

\bigskip

\noindent Proof: From (\ref{consd}), noting (\ref{tsspm}), 
Parsaval equality, we get
\bea
\label{ctd}
\|G_0 \bPsi^{n+1}\|_{l^2}^2&=&\|G_0 \bPsi^{(4)}\|_{l^2}^2
=\|G_0 \bPsi^{(3)}\|_{l^2}^2 = 
\|P\; e^{-iD \int_{t_n}^{t_{n+1}} 
f(\cdot,s)\; ds} \; P^T \; G_0\; \bPsi^{(2)}\|_{l^2}^2 \nn\\
&=&\|G_0 \bPsi^{(2)}\|_{l^2}^2 =\|G_0 \bPsi^{(1)}\|_{l^2}^2
=\|G_0 \bPsi^{n}\|_{l^2}^2, \qquad n=0,1,\cdots\; .
\eea
The conservation (\ref{consd}) is obtained from (\ref{ctd})
by induction. When $f\equiv 0$, the proof of (\ref{consdd})
follows the line of the analogous result for the linear
Schr\"{o}dinger equation by time-splitting Fourier-spectral 
approximation in \cite{Bao1,Bao4}.

\bigskip

\begin{remark}
When the definite integral $\int_{t_n}^{t_{n+1}} 
f(s)\; ds$ in (\ref{tsspm}) could not be evaluated analytically for
some very complicated function $f$, it can be evaluated numerically
using a numerical quadrature, e.g., the  Simpson's  rule
\[\int_{t_n}^{t_{n+1}}f(s)\; ds\approx \fl{k}{4}
\left[f(t_n)+4f(t_n+k/2)+
f(t_{n+1})\right], \quad j=0,\cdots,N, \ n\ge0.\]
\end{remark}

\begin{remark}
When $M=2$, $b_{11}=b_{22}=0$, $b_{12}=b_{21}=1$
 and $\hat{f}(t)=\Og \cos(\og_d\; t)$ in (\ref{vgpe}) \cite{Hall}, 
thus $f(t)=\fl{\Og}{\og_{x,1}}\cos(\og_d\; t/\og_{x,1})$ 
in (\ref{gpe2}).
After a simple computation, we can get explicitly the solution 
(\ref{sestep3})of the ODE system  (\ref{sstep3}) which will be
used in our numerical example 5 in the next section:
\beas
&&\psi_1(x,t)=\cos(g(t,t_n))\;\psi_1(x,t_n) - 
i \sin(g(t,t_n))\;\sqrt{N_2^0/N_1^0}\; \psi_2(x,t_n), \\
&&\psi_2(x,t)=- i \sin(g(t,t_n))\;\sqrt{N_1^0/N_2^0}\; \psi_1(x,t_n)
+\cos(g(t,t_n))\;\psi_2(x,t_n), \quad t_n\le t\le t_{n+1};
\eeas
where
\beas
g(t,t_n)&=&\int_{t_n}^t f(s) \;ds = \int_{t_n}^t 
\fl{\Og}{\og_{x,1}}\cos(\og_d\; s/\og_{x,1})\; ds \\
&=&\fl{\Og}{\og_d}\left[\sin(\og_d\; t/\og_{x,1})-\sin(\og_d\; t_n/\og_{x,1})
\right].
\eeas
\end{remark}

\section{Numerical results}\label{sne} 
\setcounter{equation}{0} 
 
In this section we report the coupled basis
wavefunctions with lowest energy of a two-state model  and
ground states of multi-component BEC by using the normalized 
gradient flow method, and dynamics of multi-component BEC by using
the time-splitting sine-spectral method. Furthermore we
also give a physical discussion on our numerical results.

\bigskip

\noindent {\bf Example 1} Coupled basis
wavefunctions with lowest energy of a two-state model, i.e.
we choose $\bt=100$ and $F=0.79$ in (\ref{tsmse}), (\ref{tsmve}).
We solve this problem on $[0,8]$, i.e. $R=8$ with mesh size 
$h_r=\fl{1}{64}$ and time step $k=0.1$ by using 
the BEFD discretization (\ref{tsmbe}),(\ref{tsmce}). 
Ths initial data in (\ref{tsngf4})
is chosen as
\[\phi_{s,0}(r)= \fl{1}{\pi^{1/2}}e^{-r^2/2}, \qquad
\phi_{v,0}(r)= \fl{r}{\pi^{1/2}}e^{-r^2/2}, \qquad r\ge0.\]
The steady state solution is reached when $\|\phi_s^{n+1}-\phi_s^n\|_{l^2}
+\|\phi_v^{n+1}-\phi_sv^n\|_{l^2}<\vep=10^{-6}$. Table 1 displays 
the values of $\phi_s(0)$, energies $E_s$, $E_v$ and $E$, 
chemical potentials $\mu_s$, $\mu_v$. Figure 1 shows the coupled basis
wavefunctions $\phi_{s,g}(r)$ and $\phi_{v,g}(r)$ for different 
vortex fraction $0\le n_v\le 1$, and Figure 2 shows 
surface plots of the atomic density function 
$|\psi|^2 = \left| a_s \phi_{s,g}+ a_v\phi_{v,g}e^{i\tht}\right|^2$
with $a_v=\sqrt{n_v}$ and $a_s=\sqrt{1-n_v}$ for different 
vortex fraction $0\le n_v\le 1$.

\begin{table}[htbp]
\begin{center}
\begin{tabular}{cccccccc}\hline
$n_v$ &$n_s$ &$\phi_s(0)$ &$E_s$ &$E_v$ &$E$ &$\mu_s$ &$\mu_v$ \\
0   &1 &0.2381 &3.9459  &NA &3.9459 &5.7598 &NA\\
0.1 &0.9 &0.2517 &3.8697 &5.8901 &4.0717 &5.7939 &6.9516 \\
0.3 &0.7 &0.2875 &3.7370 &5.4928 &4.2637 &5.8864 &6.6513 \\
0.5 &0.5 &0.3433 &3.6474 &5.1110 &4.3792 &6.0166 &6.4091 \\
0.7 &0.3 &0.4450 &3.6291 &4.7618 &4.4220 &6.1946 &6.2276 \\
0.9 &0.1 &0.7113 &3.7653 &4.4637 &4.3938 &6.4023 &6.0951\\
1 &0 &NA &NA &4.3689 &4.3689 &NA &6.0514 \\
    \hline
\end{tabular}
\caption{Numerical results for a two-state model in 2d in Example 1.}
\end{center}
\end{table}

From Table 1 and Figures 1\&2, we can see that, when the 
vortex state fraction 
$n_v$ increases from  $0$ to $1$, the central value of the 
symmetric state $\phi_s(0)$ and the total energy 
$E$ increase,  chemical potentials of the symmetric 
state $\mu_s$ and vortex state $\mu_v$  increases and decreases,
respectively; the atomic density function 
$|\psi|^2$ changes from ground state (cf. Fig. 2a) 
to vortex state (cf. Fig. 2f).

\bigskip

\noindent {\bf Example 2} Ground state of two-component BEC in 3d 
with dynamically stable inter-component
interaction, i.e. $a_{11}>0$, $a_{22}>0$ and $a_{11}a_{22}-a_{12}^2>0$
\cite{Jaksch}.  We choose $M=2$, 
$m=1.44\tm10^{-25}\;[kg]$, $a_{11}=a_{22}=5.45\;[nm]$,
$a_{12}=a_{21}=5.24\;[nm]$, $\og_{x,1}=\og_{y,1}=\og_{x,2}=\og_{y,2}=
10\tm 2\pi\;[1/s]$, $\og_{z,1}=\og_{z,2}=\sqrt{8}\og_{x,1}$, 
$\hat{x}_{0,1}=\hat{x}_{0,2}=\hat{y}_{0,1} =\hat{y}_{0,2}
= \hat{z}_{0,1}=\hat{z}_{0,2}=0$, $\hat{f}\equiv0$ in 
(\ref{vgpe}). Plugging these parameters into (\ref{gpe2}), we
get the dimensionless parameters 
$a_0=0.3407\tm 10^{-5}\;[m]$, $\bt_{11}=0.02010177N_1^0$, 
$\bt_{12}=0.0193272N_2^0$, $\bt_{21}=0.0193272N_1^0$, 
$\bt_{22}=0.02010177N_2^0$. We compute the ground states of
this problem in cylindrical coordinate on
$(r,z)\in \Og=[0,8]\tm[-4,4]$ with mesh size $h=h_r=h_z=\fl{1}{32}$
and time step $k=0.1$ by using the BEFD discretization for
different $N_1^0$ and $N_2^0$. Here we report the results 
for two cases:

\bigskip

\noindent Case I. $N_2^0=N_1^0$;

\noindent Case II. $N_2^0=2N_1^0$.

\bigskip

Table 2 displays the central densities $\phi_{g,1}(0,0)^2$,
$\phi_{g,2}(0,0)^2$, chemical potential $\mu_{g,1}$, $\mu_{g,2}$
and energy $E_\bt$ for case I with different $N_1^0$. 
Figure 3 shows the ground state condensate wave functions 
for case I. Furthermore Table 3 and Figure 4 
show similar results for case II.

 \begin{table}[htbp]
\begin{center}
\begin{tabular}{cccccc}\hline
$N_1^0$  &$\phi_{g,1}^2(0,0)$ &$\mu_{g,1}$  &$\phi_{g,2}^2(0,0)$ &$\mu_{g,2}$
&$E$ \\
0   &0.5496 &2.4130  &0.5496  &2.4130  &2.4130\\
100 &0.4747   &2.7664   &0.4747     &2.7664     &2.5994\\
500 &0.3548   &3.6406   &0.3548     &3.6406      &3.1161\\
1,000  &0.2969   &4.3481    &0.2969     &4.3481      &3.5650\\
3,000  &0.2170   &6.0980   &0.2170     &6.0980       &4.7258\\
6,000  &0.1765   &7.7461   &0.1765     &7.7461       &5.8504\\
10,000  &0.1513   &9.3204  &0.1513     &9.3204     &6.9388\\
20,000 &0.1226  &12.0802     &0.1226    &12.0802      &8.8655\\
    \hline
\end{tabular}
\caption{Numerical results for the ground states of two-component
BEC in 3d in example 2 for case I.}
\end{center}
\end{table}  

   \begin{table}[htbp]
\begin{center}
\begin{tabular}{cccccc}\hline
$N_1^0$  &$\phi_{g,1}^2(0,0)$ &$\mu_{g,1}$  &$\phi_{g,2}^2(0,0)$ &$\mu_{g,2}$
&$E$ \\
10   &0.5353  &2.4738   &0.5351     &2.4746     &2.4440\\ 
100  & 0.4504   &2.9062     &0.4491     &2.9122     &2.6799\\
500   &0.3225   &4.0142   &0.3193     &4.0315   &3.3582\\
1,000  &0.2679    &4.8777   &0.2637     &4.9029    &3.9218\\
3,000  &0.1963   &6.9773    &0.1902     &7.0200     &5.3429\\ 
6,000   &0.1610   &8.9353   &0.1536     &8.9931    &6.6984\\
10,000  &0.1392    &10.7975  &0.1309    &10.8692   &8.0013\\
20,000   &0.1145  &14.0520    &0.1053    &14.1471  &10.2956\\
    \hline
\end{tabular}
\caption{Numerical results for the ground states of two-component
BEC in 3d in example 2 for case II.}
\end{center}
\end{table}

From Tables 2\&3 and Figures 3\&4, we can see that, when the number
of particles of the two component are the same, i.e. 
$N_1^0=N_1^0$, then the ground state density functions for the
two components are equal to each other, i.e. 
$\phi_{g,1}^2=\phi_{g,2}^2$ due to the same parameters used for the
two components; where when  $N_1^0\ne N_2^0$, then
$\phi_{g,1}^2\ne \phi_{g,2}^2$.
In the two cases, when the number
of particles in the first condensate $N_1^0$ increases, 
the central value of the density functions 
$\phi_{g,1}^2(0,0)$ and $\phi_{g,1}^2(0,0)$ decrease, but the 
 the total energy 
$E$, and the  chemical potentials  $\mu_{g,1}$ and $\mu_{g,2}$
increase.

\bigskip

\noindent {\bf Example 3} Ground state of two-component BEC in 3d 
with dynamically unstable inter-component
interaction, i.e. $a_{11}>0$, $a_{22}>0$ and $a_{11}a_{22}-a_{12}^2<0$
\cite{Hall}.  We choose $M=2$, 
$m=1.44\tm10^{-25}\;[kg]$, $a_{12}=a_{21}=55.3\AA=5.53\;[nm]$, 
$a_{11}=1.03a_{12}=5.6959\;[nm]$,  $a_{22}=0.97a_{12}=5.3641\;[nm]$,
 $\og_{z,1}=\og_{z,2}=47\tm 2\pi\;[1/s]$, 
$\og_{x,1}=\og_{y,1}=\og_{x,2}=\og_{y,2}=\og_{z,1}/\sqrt{8}$,  
$\hat{x}_{0,1}=\hat{x}_{0,2}=\hat{y}_{0,1} =\hat{y}_{0,2}=0$, 
$\hat{f}\equiv0$ in 
(\ref{vgpe}). Plugging these parameters into (\ref{gpe2}), we
get the dimensionless parameters 
$a_0=0.2643\tm 10^{-5}\;[m]$, $\bt_{11}=0.02708165N_1^0$, 
$\bt_{12}=0.02629286N_2^0$, $\bt_{21}=0.02629286N_1^0$, 
$\bt_{22}=0.02550407N_2^0$. We compute the ground states of
this problem in cylindrical coordinate on
$(r,z)\in \Og=[0,16]\tm[-12,12]$ with mesh size $h=h_r=h_z=\fl{1}{16}$
and time step $k=0.1$ by using the BEFD discretization for
different $N_1^0$ and $N_2^0$. Here we report the results 
for three cases:

\bigskip

\noindent Case I. $\hat{z}_{0,1}=\hat{z}_{0,2}=0$, $N^0=N_1^0+N_2^0=
1,000,000$. Varying the ratio between $N_1^0$ and $N_2^0$;

\noindent Case II. $N_2^0=N_1^0=500,000$. $\hat{z}_{0,1}=-\hat{z}_{0,2}\ne0$.
Varying $\hat{z}_{0,1}$;

\noindent Case II. $\hat{z}_{0,1}=-\hat{z}_{0,2}=0.15a_0$, $N^0=N_1^0+N_2^0=
1,000,000$. Varying the ratio between $N_1^0$ and $N_2^0$.

\bigskip

Table 4 displays the central densities $\phi_{g,1}(0,0)^2$,
$\phi_{g,2}(0,0)^2$, chemical potential $\mu_{g,1}$, $\mu_{g,2}$
and energy $E_\bt$ for case I with different $N_1^0$. 
Figure 5 shows the ground state condensate wave functions 
for case I. Furthermore Table 5 and Figure 6 
show similar results for case II and Table 6 and Figure 7
for case III.

 \begin{table}[htbp]
\begin{center}
\begin{tabular}{ccccccc}\hline
$N_1^0$ &$N_2^0$ &$\phi_{g,1}^2(0,0)$ &$\mu_{g,1}$  
  &$\phi_{g,2}^2(0,0)$ &$\mu_{g,2}$  &$E$ \\
100,000 &900,000   &0.0007  &47.8143    &0.0453    &47.2278   &33.8714\\
300,000 &700,000   &0.0025  &47.9650   &0.0514    &47.2456  &34.0028\\
500,000 &500,000  &0.0063   &48.0880  &0.0605    &47.2631    &34.1575\\
700,000 &300,000   &0.0129   &48.1999  &0.0759    &47.2779   &34.3323\\
900,000 &100,000  &0.0270   &48.3077  &0.1082   &47.2873  &34.5266\\
    \hline
\end{tabular}
\caption{Numerical results for the ground states of two-component
BEC in 3d in example 3 for case I.}
\end{center}
\end{table}

 \begin{table}[htbp]
\begin{center}
\begin{tabular}{cccccc}\hline
$\hat{z}_{0,1}/a_0$  &$\phi_{g,1}^2(0,0)$ &$\mu_{g,1}$  
  &$\phi_{g,2}^2(0,0)$ &$\mu_{g,2}$  &$E$ \\
0.01    &0.0092   &48.0611    &0.0601    &47.2363     &34.1375\\
0.1  &0.0307    &47.3400    &0.0513    &46.4531    &33.5336\\
0.5   &0.0370   &43.9392  &0.0425    &43.0278    &30.8042\\
2.0    &0.0277    &37.3581   &0.0271    &36.4870     &26.2717 \\
4.0   &0.0001   &36.7085   &0.0000    &35.8441   &26.0203\\
    \hline
\end{tabular}
\caption{Numerical results for the ground states of two-component
BEC in 3d in example 3 for case II.}
\end{center}
\end{table}

\begin{table}[htbp]
\begin{center}
\begin{tabular}{ccccccc}\hline
$N_1^0$ &$N_2^0$ &$\phi_{g,1}^2(0,0)$ &$\mu_{g,1}$  
  &$\phi_{g,2}^2(0,0)$ &$\mu_{g,2}$  &$E$ \\
100,000 &900,000  &0.0023   &44.3215   &0.0452    &47.0379   &33.4594\\
300,000 &700,000    &0.0129   &45.9058  &0.0502    &46.5788    &33.1356\\
500,000 &500,000  &0.0335   &46.8865  &0.0489    &45.9868   &33.1615\\
700,000 &300,000    &0.0463   &47.6052    &0.0285    &45.1929    &33.4916\\
900,000 &100,000   &0.0443   &48.1503   &0.0072    &43.9508    &34.1429\\
    \hline
\end{tabular}
\caption{Numerical results for the ground states of two-component
BEC in 3d in example 3 for case III.}
\end{center}
\end{table}

From Tables 4,5\&6 and Figures 5,6\&7, we can have the following 
observations: (i). In case I, the trap potentials for the two components
are the same, when the fraction of the number of particles 
in the first component, i.e. $N_1^0/N^0$, increases, 
the energy $E$ increases, and the chemical potentials
for the two components, $\mu_{g,1}$ and $\mu_{g,2}$, increases
and decreases, respectively. The reason of this is due to 
$a_{11}>a_{22}$. Furthermore, we observe a crater in the
density function of the first component , 
corresponding to a shell in which the second component
reside (cf. Fig. 5e\&f). This confirms the experimental 
results (cf. Fig. 1 in \cite{Hall}).  (ii). In case II, when the distance 
between the center of the trap potentials for the two
components, i.e. $\hat{z}_{0,1}-\hat{z}_{0,2}$, increases, 
the energy $E$, chemical potentials for the two components,
 $\mu_{g,1}$ and $\mu_{g,2}$, decrease. Furthermore, the bigger
the distance, the more separation in the density functions
of the two components (cf. Fig.6). (iii). The above observation 
(i) also holds for case III except that the crater in the density
function for the first component almost disappears (cf. Fig. 7). 
This is due to  the separation of the centers of the trap potentials for the
two components.

\bigskip

\noindent {\bf Example 4} Dynamics of two-component BEC in 3d 
with dynamically unstable inter-component
interaction, i.e. $a_{11}>0$, $a_{22}>0$ and $a_{11}a_{22}-a_{12}^2<0$
\cite{Hall}.  We choose $M=2$, 
$m=1.44\tm10^{-25}\;[kg]$, $a_{12}=a_{21}=55.3\AA=5.53\;[nm]$, 
$a_{11}=1.03a_{12}=5.6959\;[nm]$,  $a_{22}=0.97a_{12}=5.3641\;[nm]$,
 $\og_{z,1}=\og_{z,2}=47\tm 2\pi\;[1/s]$, 
$\og_{x,1}=\og_{y,1}=\og_{x,2}=\og_{y,2}=\og_{z,1}/\sqrt{8}$,  
$\hat{x}_{0,1}=\hat{x}_{0,2}=\hat{y}_{0,1} =\hat{y}_{0,2}=0$, 
$\hat{z}_{0,1}=-\hat{z}_{0,2}=0.15a_0$,
$\hat{f}(\bx,t)=\Og \cos(\og_d\;t)$ in (\ref{vgpe}). 
We start the simulation with the initial data chosen 
as the ground state of (\ref{gpe2}) computed by setting  $f\equiv0$ and
using BEFD discretization. We take $\og_d=6.5\tm 2\pi\;[1/s]$, 
$N_1^0=N_2^0=500,000$,
and solve this problem on the box $[-16,16]\tm[-16,16]\tm[-8,8]$  
with mesh sizes $h_x=h_y =1/4$,  $h_z=1/8$ and time step $k=0.0002$.
Figure 8 shows the time evolution of the mean of the density functions
for the two components, $\|\psi_1\|^2$, $\|\psi_2\|^2$ (noticing
that the number of particles in the two components are 
$N_1^0\|\psi_1\|^2$, $N_1^0\|\psi_2\|^2$, respectively) 
for $\Og=6.5\tm2\pi\;[1/s]$,  $65\tm2\pi\;[1/s]$ and $650\tm2\pi\;[1/s]$.
Furthermore Figure 9 displays the time evolution of the
density functions for the two components for $\Og=65\tm2\pi\;[1/s]$.

  \bigskip

The general form of time evolution on the number of particles
in the two components is similar for different external driven field
frequency $\Og$. When $\Og$ is small, the number of particles in the
first component, i.e. $N_1^0 \|\psi_1\|^2$, increases, attains
its peak and then decreases; where the pattern for  $N_2^0 \|\psi_2\|^2$
is opposite (cf. Fig. 8a), 
which is due to the total number of particles in the 
two components are conserved. When $\Og$ becomes bigger, 
the pattern of $N_1^0 \|\psi_1\|^2$ oscillates for some time 
period,  attains its absolute peak, and then oscillates again
(cf. Fig. 8b). Initially the density functions for the two 
components are well separated (cf. Fig. 9 first row), 
around at time $t=3.4$, the number of particles
in the first component attains its maximum and a bigger 
condensate (with approximately 52\% bigger in term of the number
of particles for the first component than that initially at time $t=0$) 
is generated (cf. Fig. 8b\&9). 
  When $\Og$ becomes very big,  similar pattern of $N_1^0 \|\psi_1\|^2$ 
is observed. In fact, the bigger $\Og$ is, the faster 
oscillation in the pattern of the number of particles in the 
condensates (cf. Fig. 8a,b\&c).

\section{Conclusions}\label{ss} 
\setcounter{equation}{0} 

The ground states and dynamics of multi-component Bose-Einstein 
condensates at temperature $T$ much smaller 
than the critical condensate temperature $T_c$ are 
studied numerically by using the time-independent vector Gross-Pitaevskii
equations (VGPRs) and time-dependent VGPEs
with (or without) an external driven field, respectively. 
We started with the 3d VGPEs, scale it to obtain a dimensionless VGPEs,
and showed how to approximately reduce it to a 2d VGPEs and a 1d VGPEs in
certain limits. We provided the approximate ground
state solution of the VGPEs in the (very) weakly interacting 
condensates. Then, most importantly,
we presented  a normalized gradient flow (NGF) to compute 
ground states of multi-component BEC, 
proved energy diminishing of the NGF which 
provides a mathematical justification, discretized 
it by the backward Euler finite difference (BEFD) which 
is monotone in linear and nonlinear cases
and preserves energy diminishing property in linear case;
as well as  we used a time-splitting 
sine-spectral method (TSSP) to discretize the 
time-dependent VGPEs with an external driven field for computing 
dynamics of multi-component BEC. The merit of the
TSSP for VGPEs is that it is explicit, unconditionally stable,
easy to program, less memory requirement, 
time reversible and time transverse invariant if the VGPEs is, 
`good' resolution in the semiclassical regime, 
of spectral order accuracy in space and second order accuracy in time, 
and conserves the total particle number in the discretized level. 
Extensive numerical examples in 3d for ground states and dynamics 
of multi-component BEC are presented to demonstrate the power of the 
numerical methods. Finally, we want to point out
that equations very similar to the VGPEs are also
encountered in nonlinear optics. In the future we plan
to apply the powerful numerical methods to study 
vortex states and their dynamical stability in 
multi-component Bose-Einstein condensates.

\bigskip 
 
\begin{center} 
{\large \bf Acknowledgment} 
\end{center} 
The author acknowledges support  by the National University of Singapore 
 grant No. R-151-000-027-112 and helpful discussions with D. Jaksch and
P. Markowich. This work was also supported in part
by the WITTGENSTEIN-AWARD of P. Markowich 
 which is funded by the Austrian National Science 
Foundation FWF.


\newpage 
 
\begin{figure}[htb] 
\centerline{a) \psfig{figure=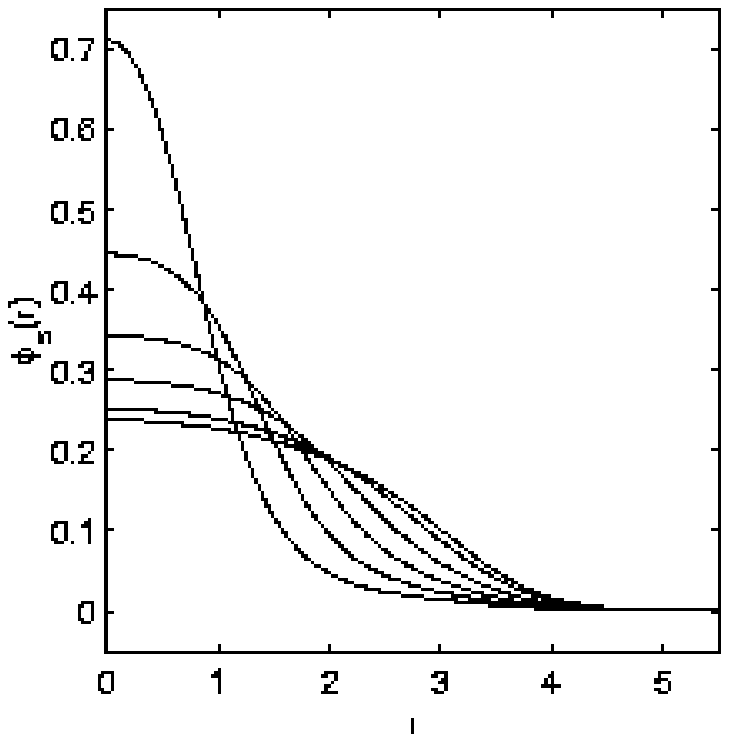,height=8cm,width=10cm,angle=0}} 
 \centerline{b) \psfig{figure=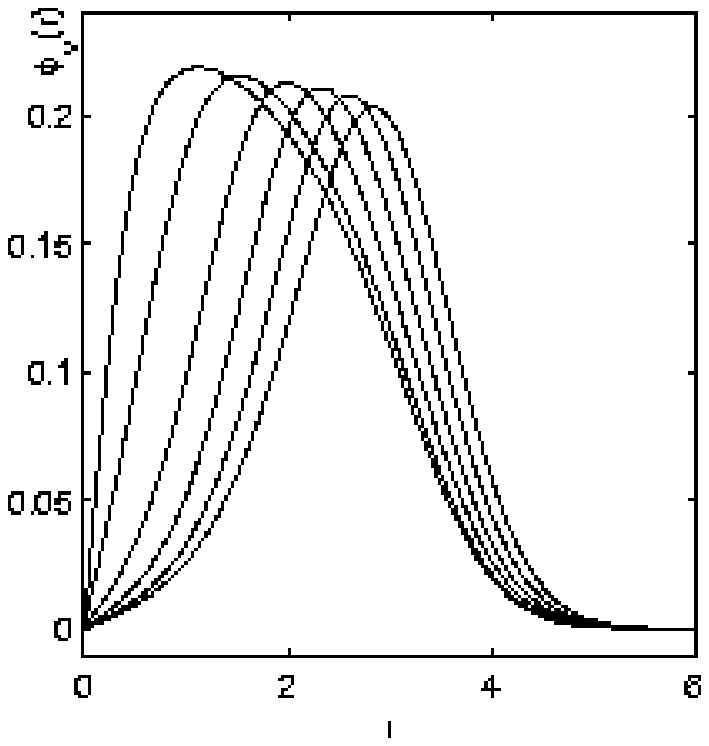,height=8cm,width=10cm,angle=0}} 
Figure 1:  The coupled basis
wavefunctions  for a two-state model in example 1
for different  vortex fraction $n_v=0, 0.1, 0.3, 0.5, 0.7, 0.9, 1$
(in the order of decreasing peak). 
a). Symmetric 
state $\phi_{s,g}(r)$,  b). Vortex state  $\phi_{v,g}(r)$.
 \end{figure}

\begin{figure}[htb] 
\centerline{a).\psfig{figure=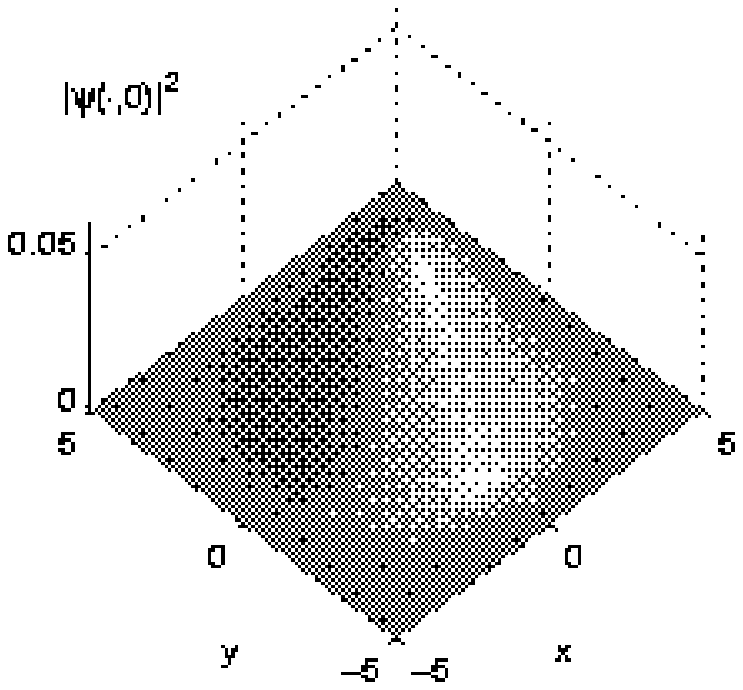,height=6.5cm,width=6.5cm,angle=0}
b).\psfig{figure=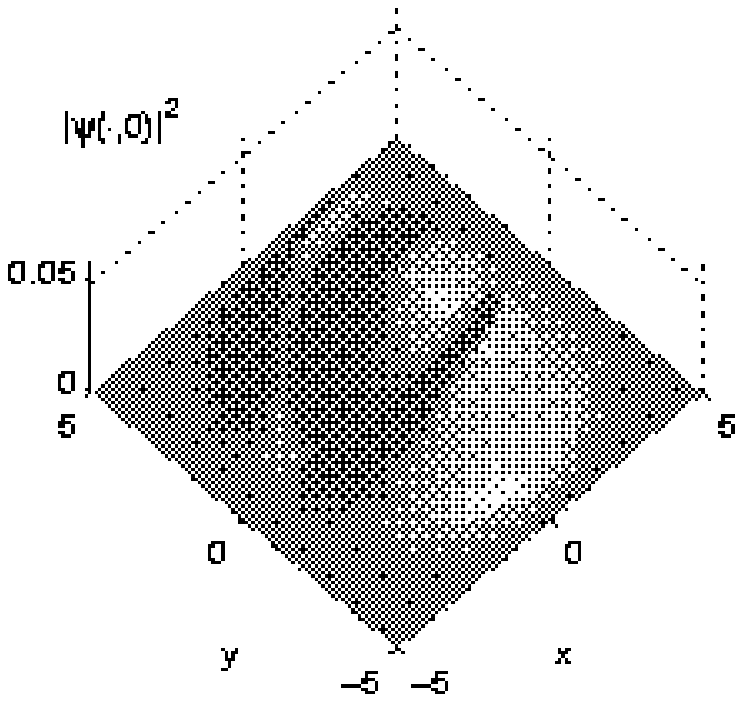,height=6.5cm,width=6.5cm,angle=0}} 
 \centerline{c).\psfig{figure=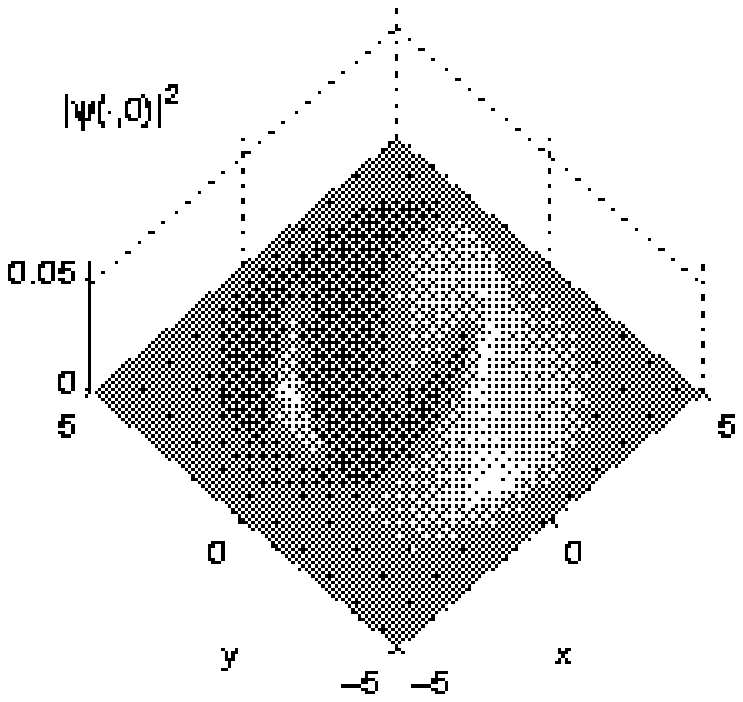,height=6.5cm,width=6.5cm,angle=0}
d).\psfig{figure=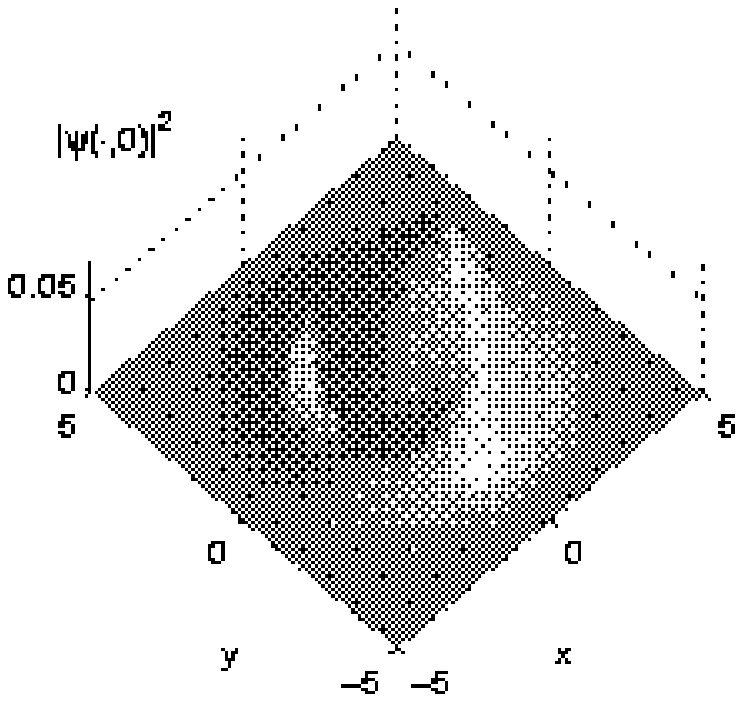,height=6.5cm,width=6.5cm,angle=0}} 
\centerline{e).\psfig{figure=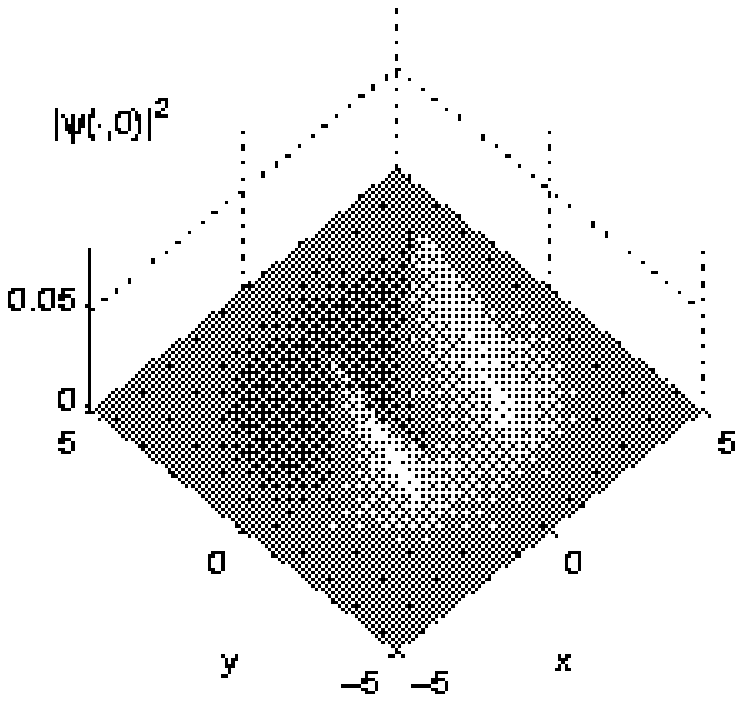,height=6.5cm,width=6.5cm,angle=0}
f).\psfig{figure=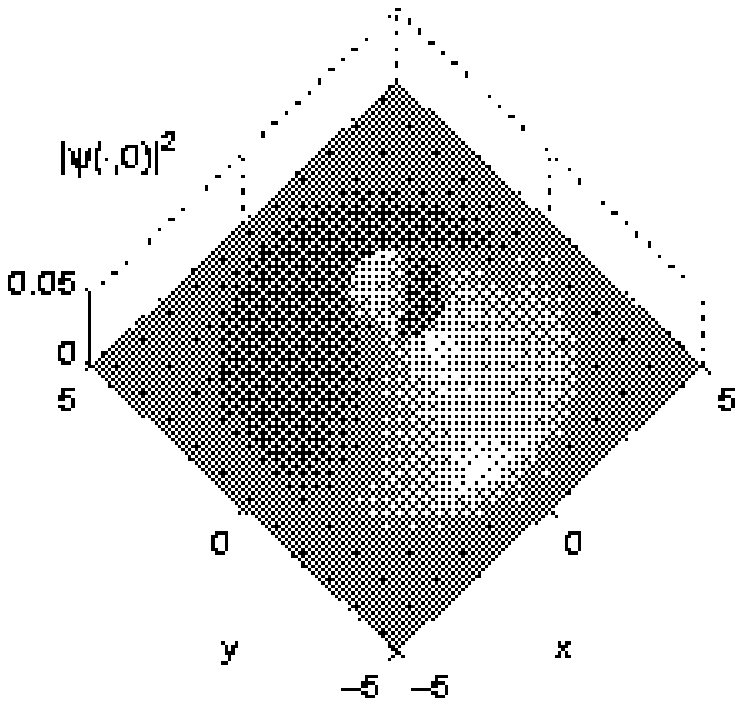,height=6.5cm,width=6.5cm,angle=0}} 

Figure 2:  Surface plot of the atomic density function 
$|\psi|^2$  for different 
vortex fraction $0\le n_v\le 1$.   a). $n_v=0$, b). $n_v=0.1$,
c). $n_v=0.3$, d). $n_v=0.5$, e). $n_v=0.9$, f). $n_v=1.0$. 
\end{figure}

\begin{figure}[htb] 
\centerline{a).\psfig{figure=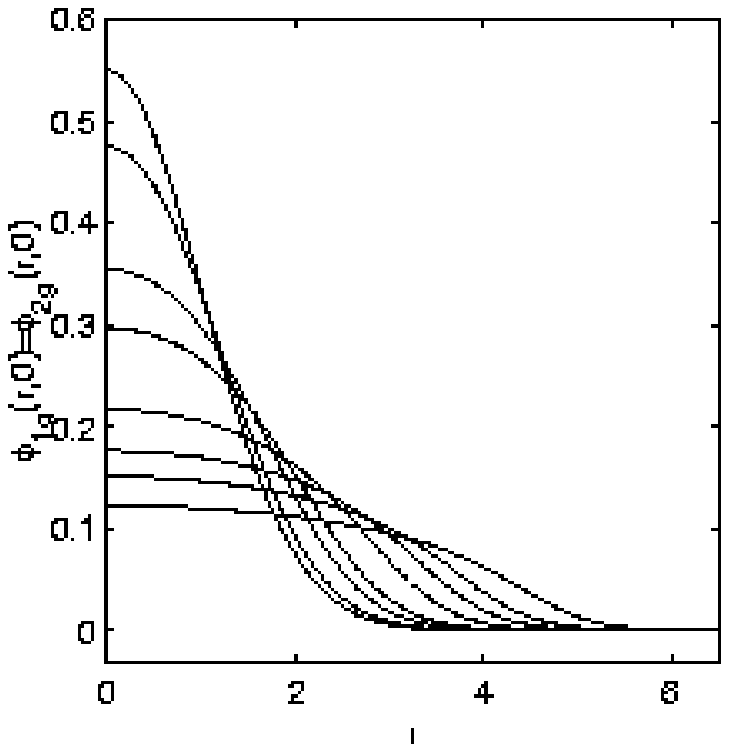,height=6.5cm,width=6.5cm,angle=0}
b).\psfig{figure=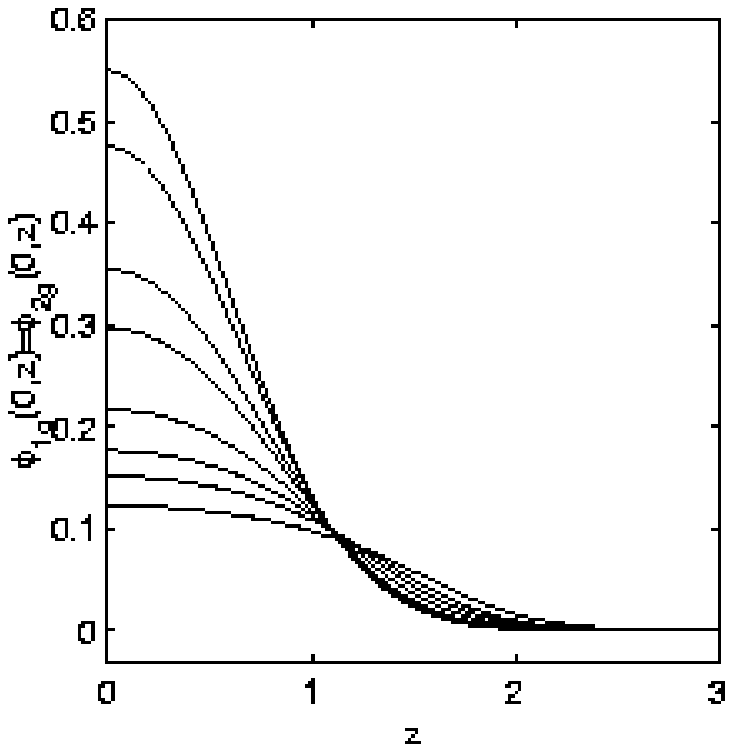,height=6.5cm,width=6.5cm,angle=0}} 
 \centerline{c).\psfig{figure=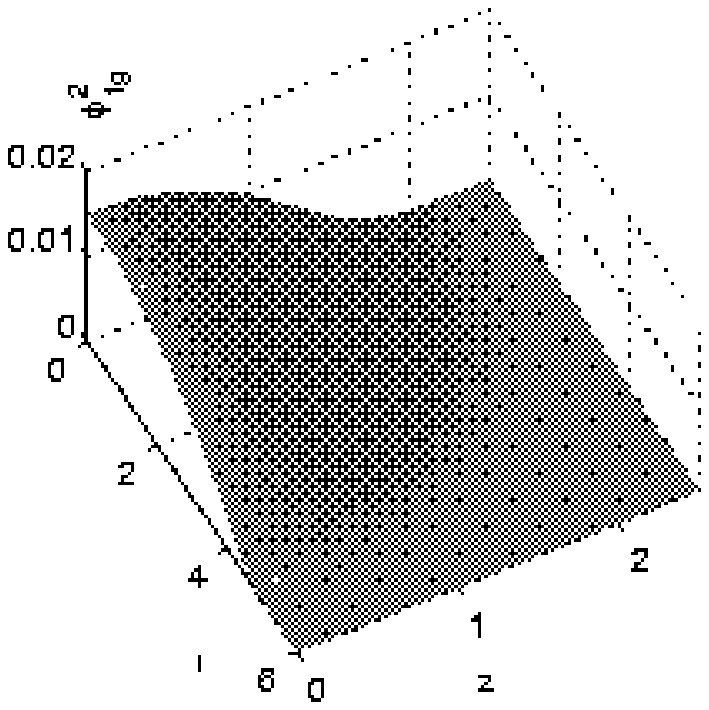,height=9cm,width=9cm,angle=0}}

Figure 3:  Ground state solution in 3d with cylindrical symmetry 
in example 2 for case I. Condensate wave function on two
lines for $N_1^0=0, 100, 500, 1000, 3000, 6000, 10000, 20000$ 
(in the order of decreasing peak): a). On the line $z=0$
$\phi_{g,1}(r,0)=\phi_{g,2}(r,0)$; b). On the line $r=0$
$\phi_{g,1}(0,z)=\phi_{g,2}(0,z)$. 
c). Surface plot of the condensate  density function 
$|\phi_{g,1}|^2=|\phi_{g,2}|^2$  for  $N_1^0=20000$. 

\end{figure}

 \begin{figure}[htb] 
\centerline{a).\psfig{figure=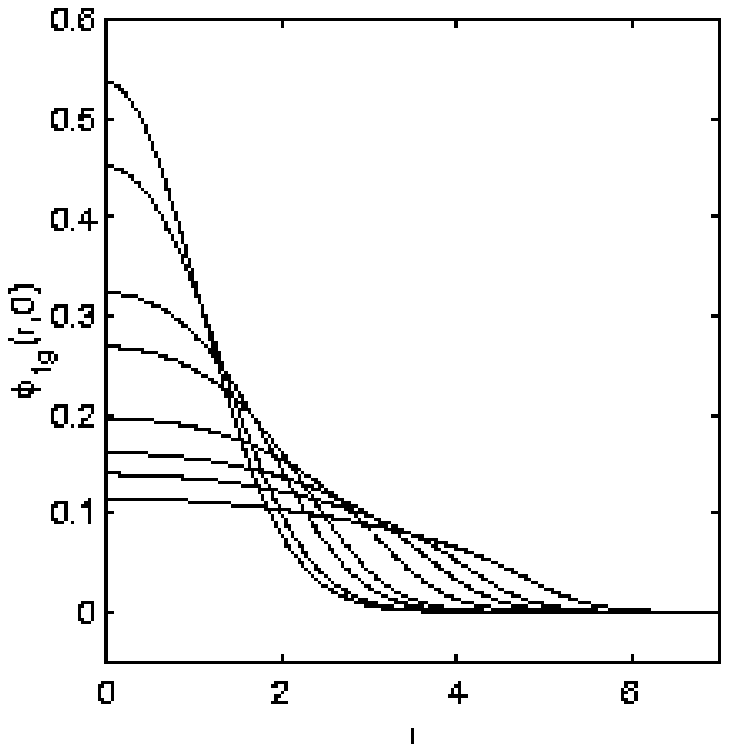,height=6.5cm,width=6.5cm,angle=0}
b).\psfig{figure=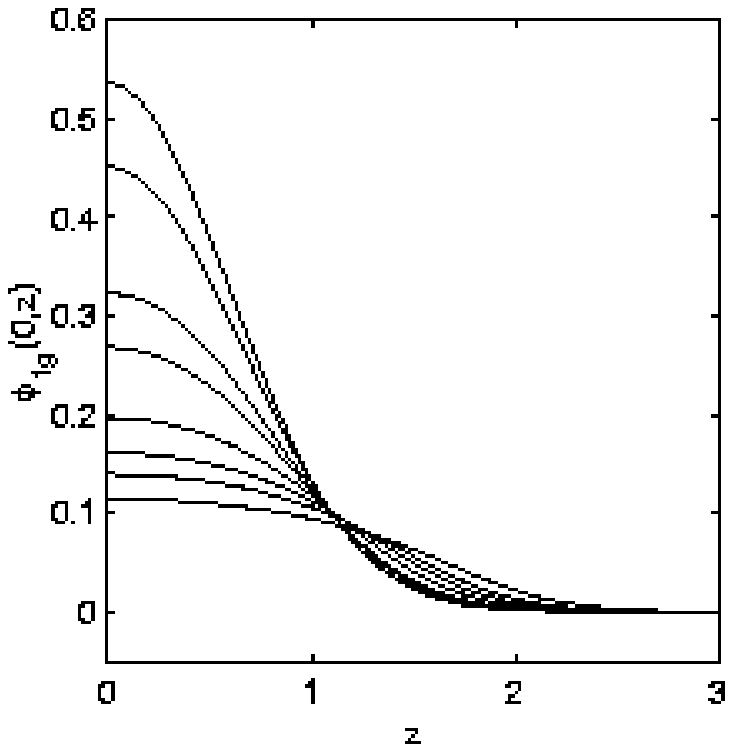,height=6.5cm,width=6.5cm,angle=0}} 
\centerline{c).\psfig{figure=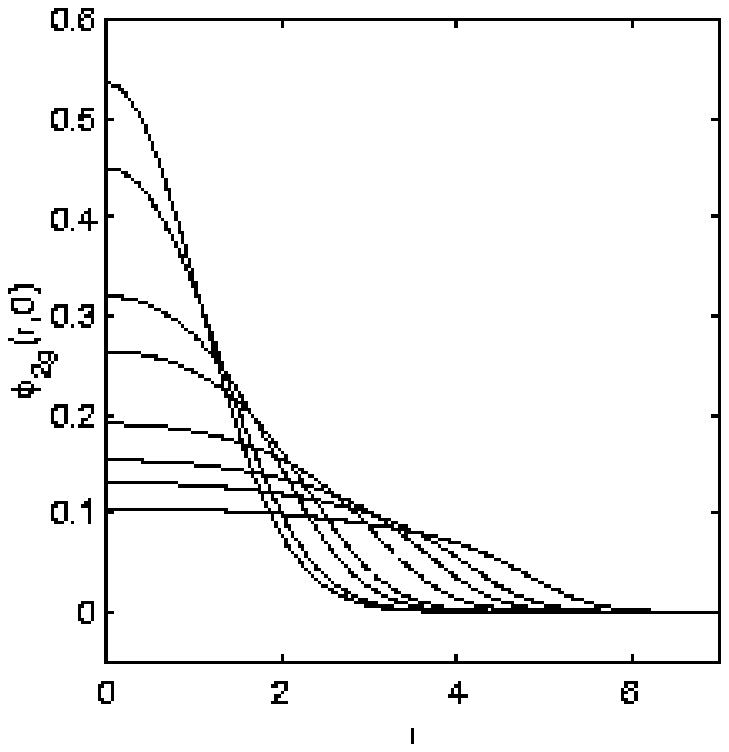,height=6.5cm,width=6.5cm,angle=0}
d).\psfig{figure=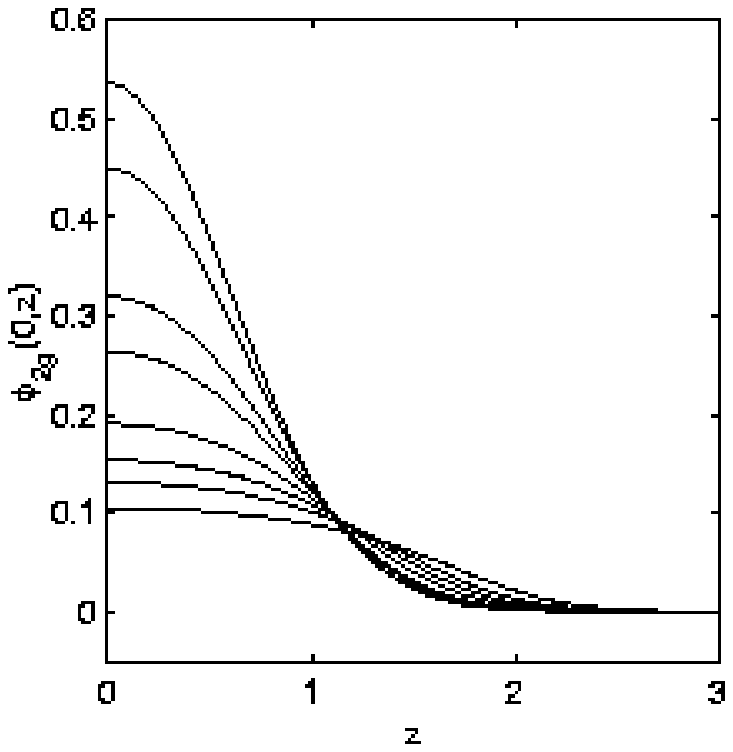,height=6.5cm,width=6.5cm,angle=0}} 
 \centerline{e).\psfig{figure=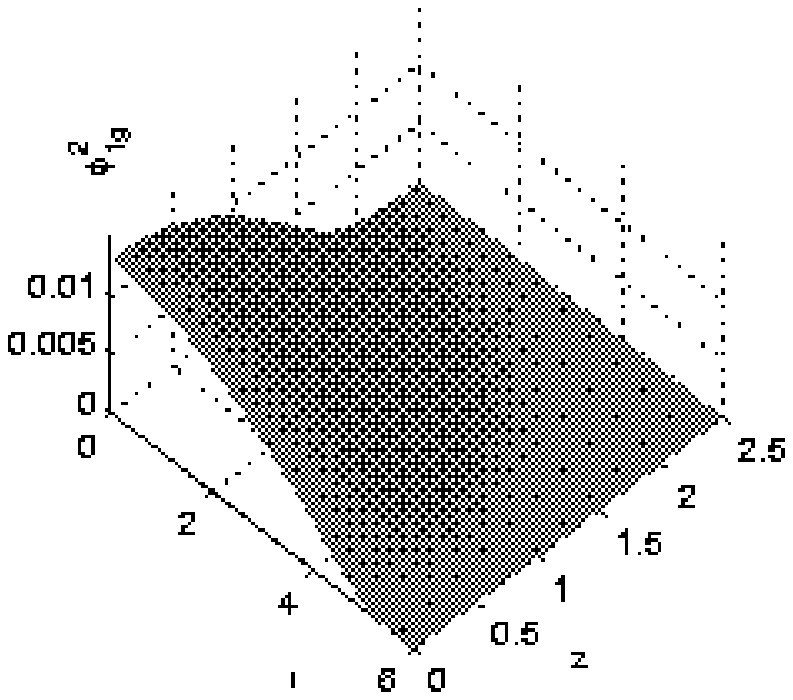,height=6.5cm,width=6.5cm,angle=0}
f).\psfig{figure=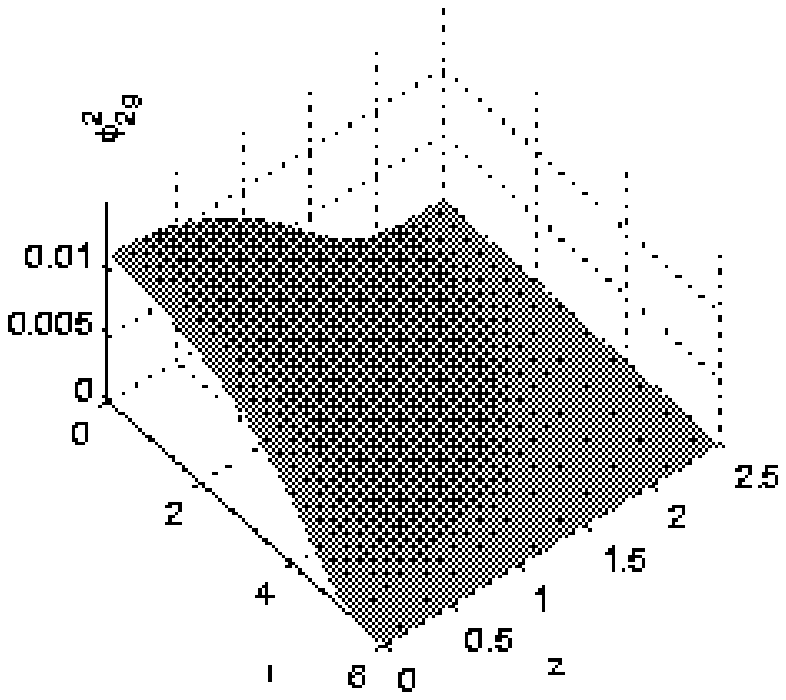,height=6.5cm,width=6.5cm,angle=0}}

Figure 4:  Ground state solution in 3d with cylindrical symmetry 
in example 2 for case II. Condensate wave function on two
lines for $N_1^0=10, 100, 500, 1000, 3000, 6000, 10000, 20000$ 
(in the order of decreasing peak). On the line $z=0$: a).  
$\phi_{g,1}(r,0)$; c). $\phi_{g,2}(r,0)$. On the line $r=0$:
 b). $\phi_{g,1}(0,z)$; d). $\phi_{g,2}(0,z)$. 
Surface plot of the condensate  density functions for  $N_1^0=20000$: 
e). $|\phi_{g,1}|^2$; f). $|\phi_{g,2}|^2$.   

\end{figure}

\clearpage

 \begin{figure}[htb] 
\centerline{a).\psfig{figure=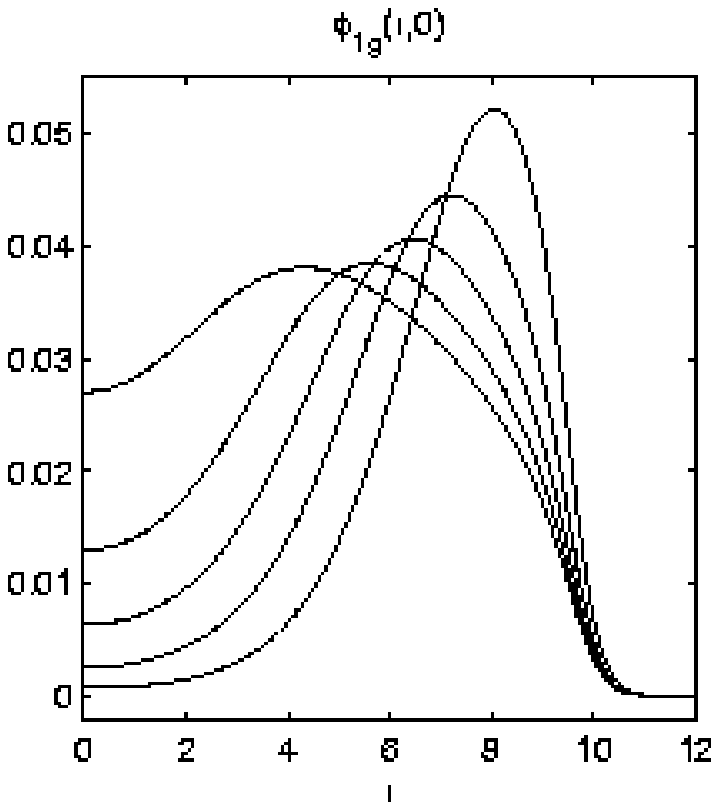,height=6.5cm,width=6.5cm,angle=0}
b).\psfig{figure=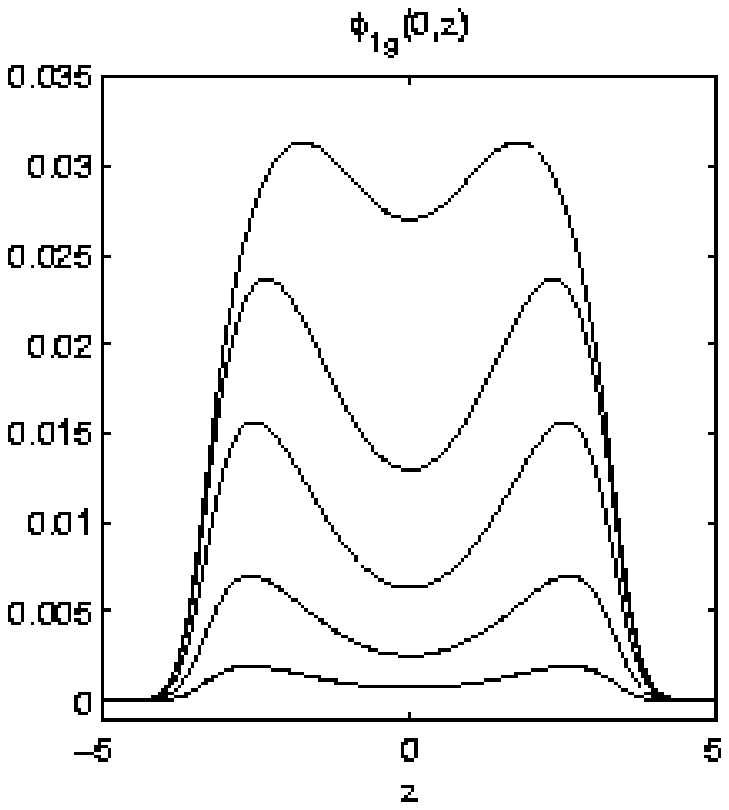,height=6.5cm,width=6.5cm,angle=0}} 
\centerline{c).\psfig{figure=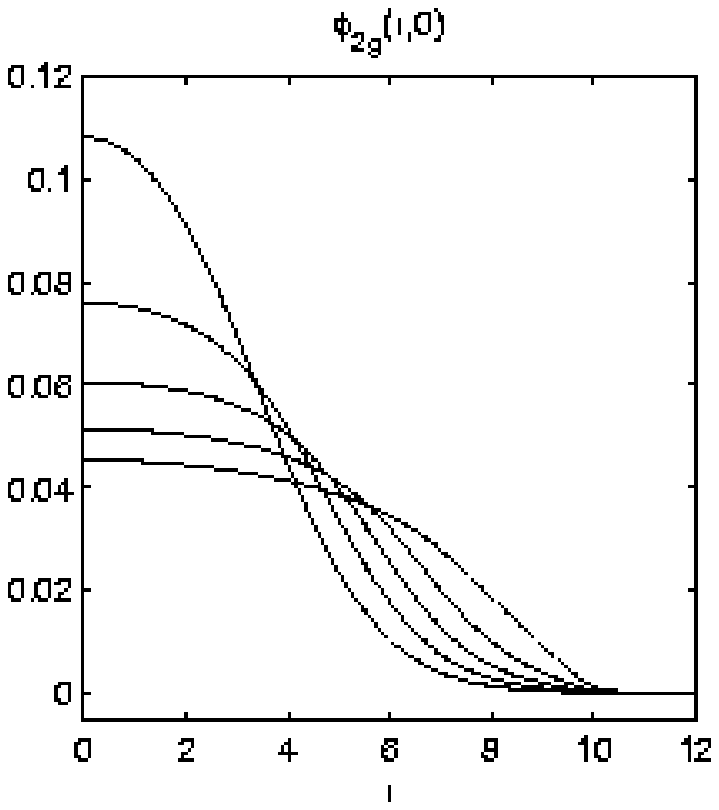,height=6.5cm,width=6.5cm,angle=0}
d).\psfig{figure=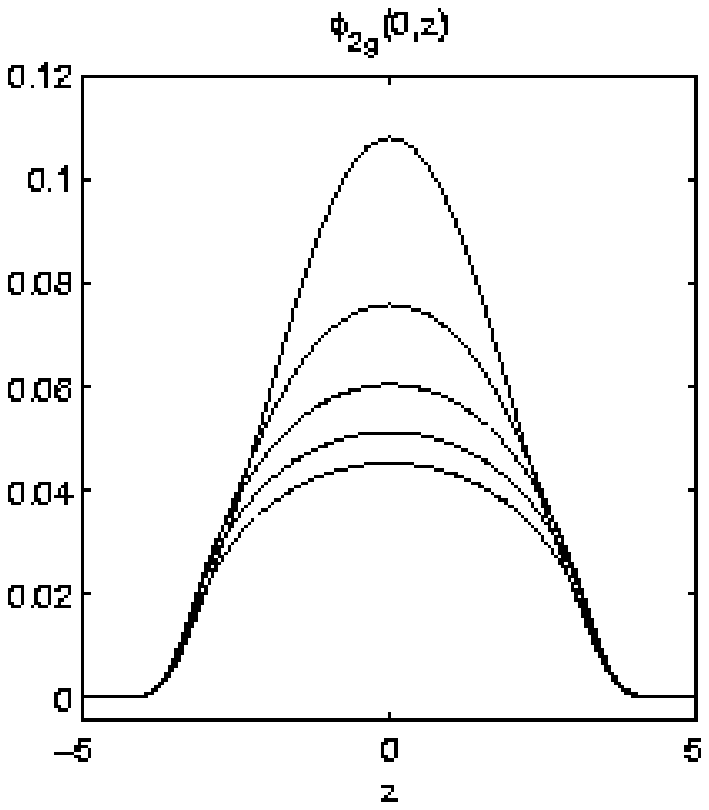,height=6.5cm,width=6.5cm,angle=0}} 
 \centerline{e).\psfig{figure=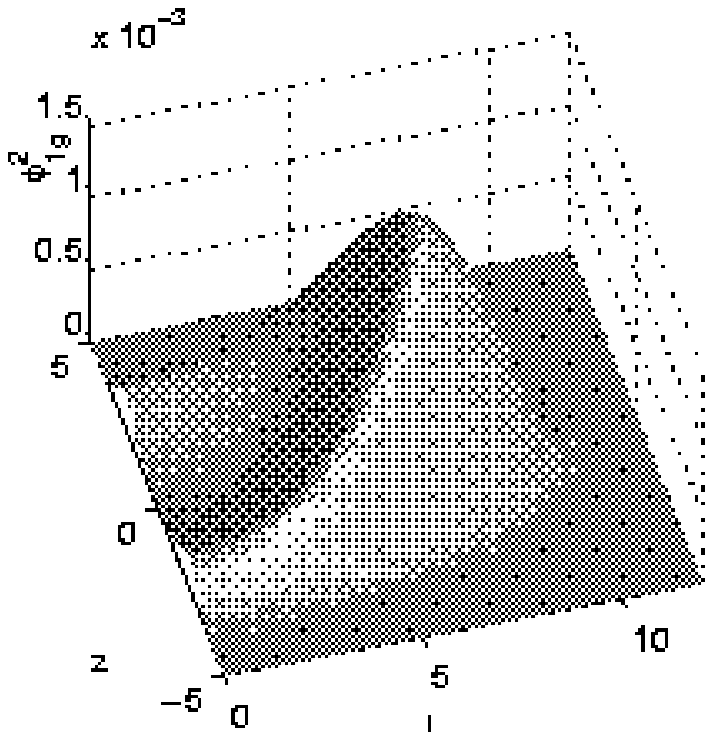,height=6.5cm,width=6.5cm,angle=0}
f).\psfig{figure=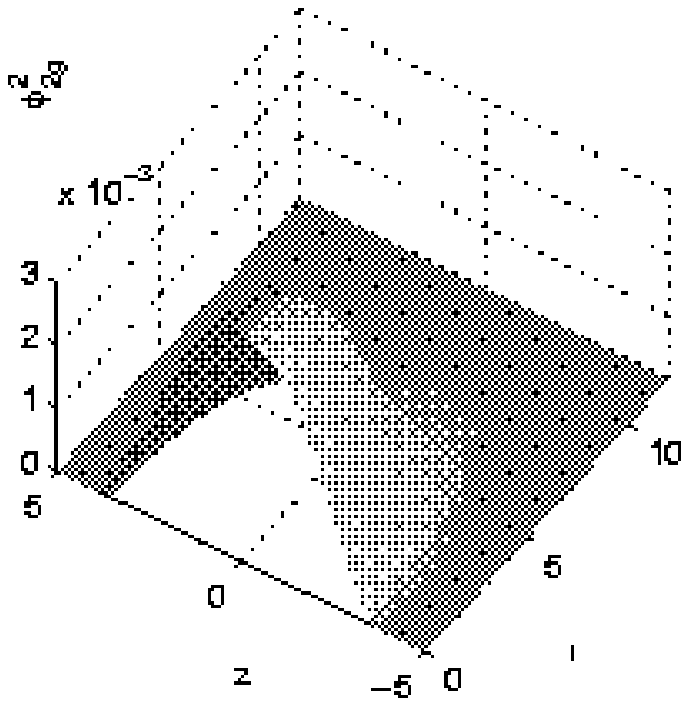,height=6.5cm,width=6.5cm,angle=0}}

Figure 5:  Ground state solution in 3d with cylindrical symmetry 
in example 3 for case I. Condensate wave function on two
lines for $N_1^0/N^0=0.1,0.3,0.5,0.7,0.9$ 
(in the order of increasing at the origin). On the line $z=0$: a).  
$\phi_{g,1}(r,0)$; c).$\phi_{g,2}(r,0)$. On the line $r=0$:
 b). $\phi_{g,1}(0,z)$; d). $\phi_{g,2}(0,z)$. 
Surface plot of the condensate  density functions for  $N_1^0=N_2^0=500,000$: 
e). $|\phi_{g,1}|^2$; f). $|\phi_{g,2}|^2$.   

\end{figure}

\begin{figure}[htb] 
\centerline{a).\psfig{figure=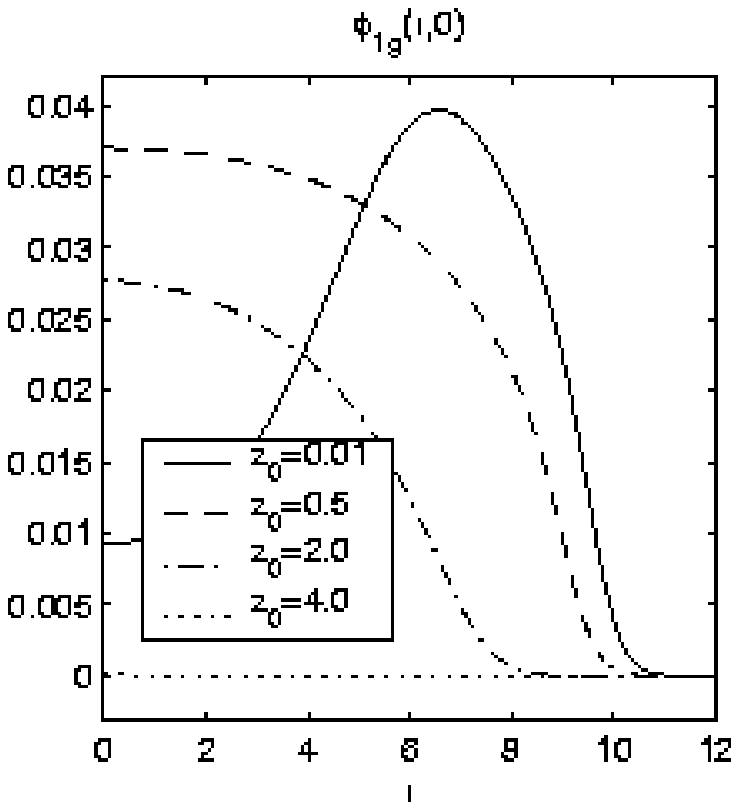,height=6.5cm,width=6.5cm,angle=0}
b).\psfig{figure=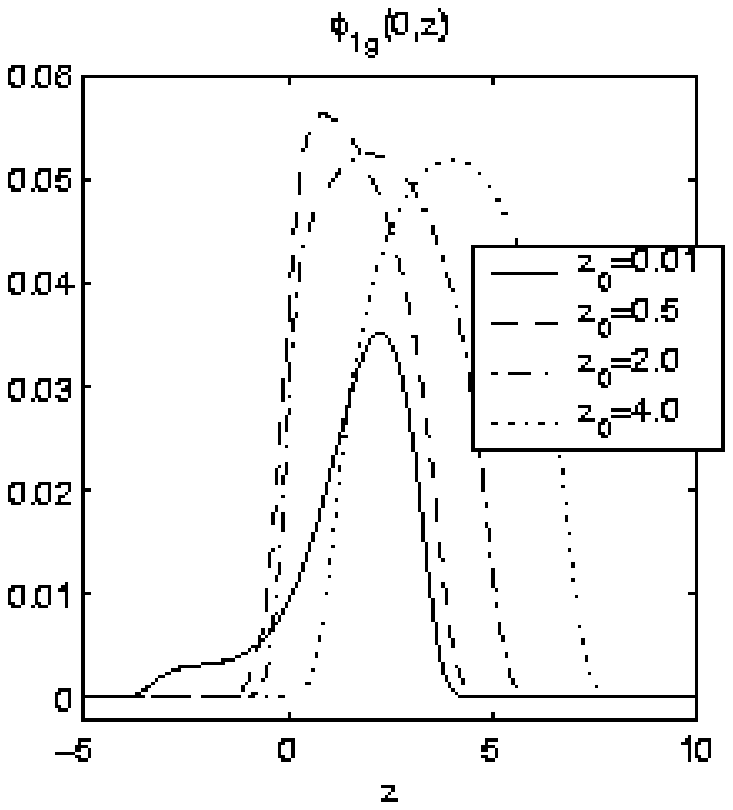,height=6.5cm,width=6.5cm,angle=0}} 
\centerline{c).\psfig{figure=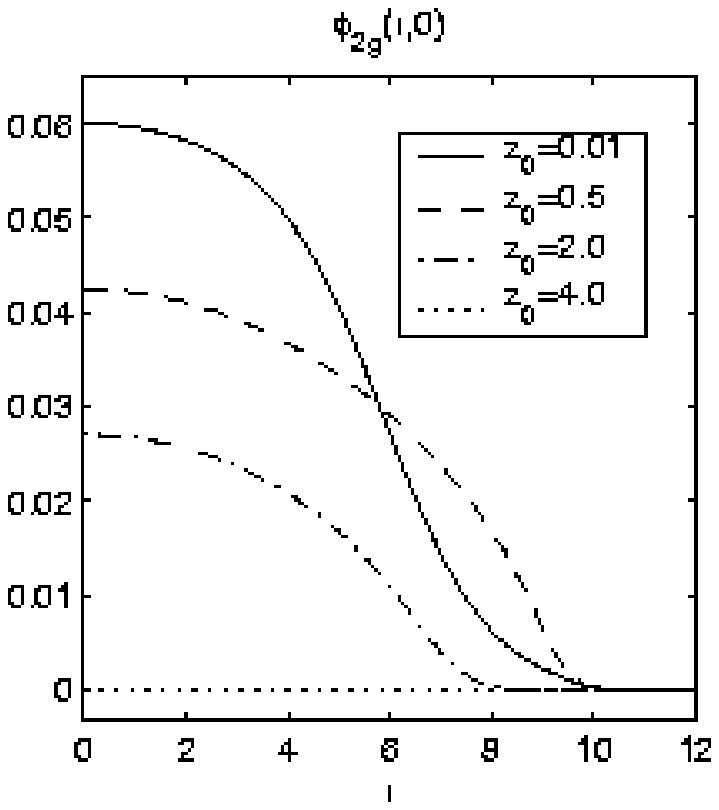,height=6.5cm,width=6.5cm,angle=0}
d).\psfig{figure=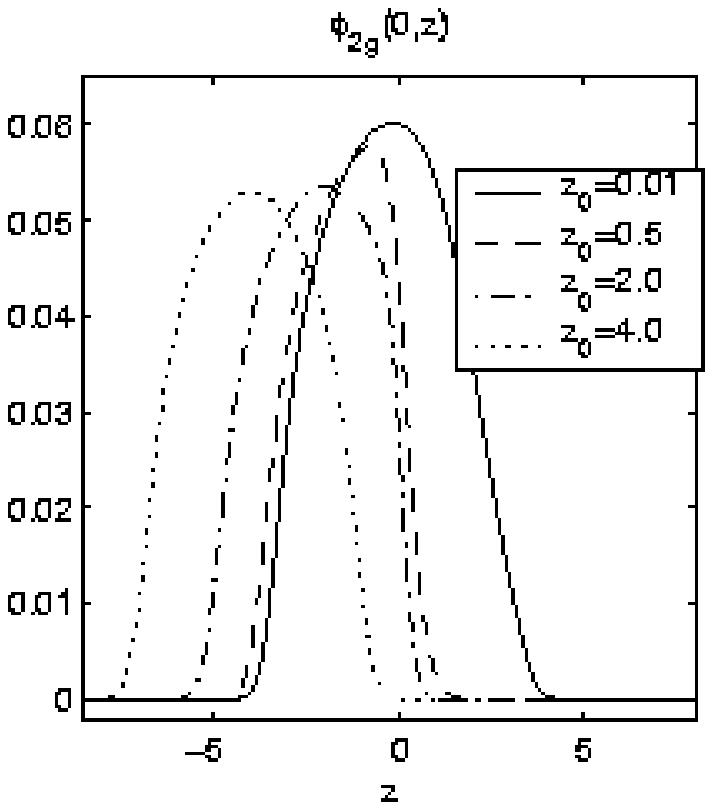,height=6.5cm,width=6.5cm,angle=0}} 
 \centerline{e).\psfig{figure=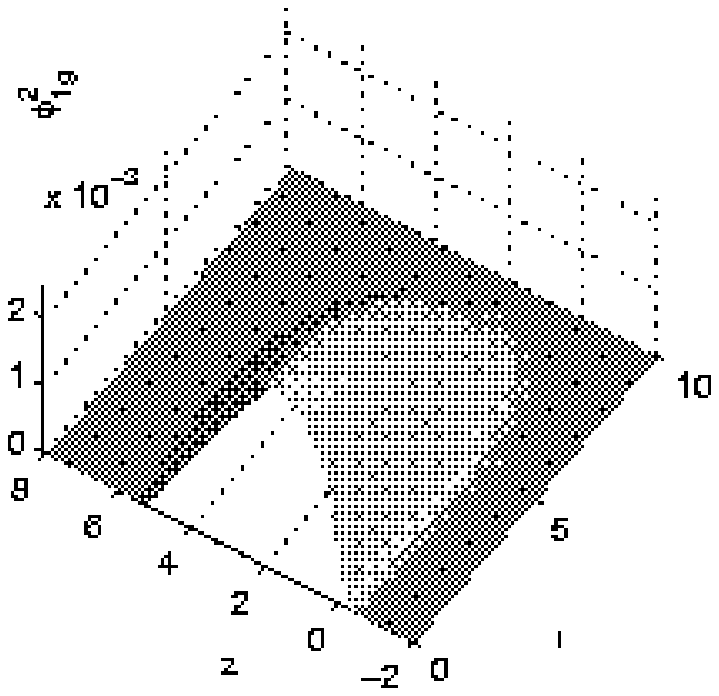,height=6.5cm,width=6.5cm,angle=0}
f).\psfig{figure=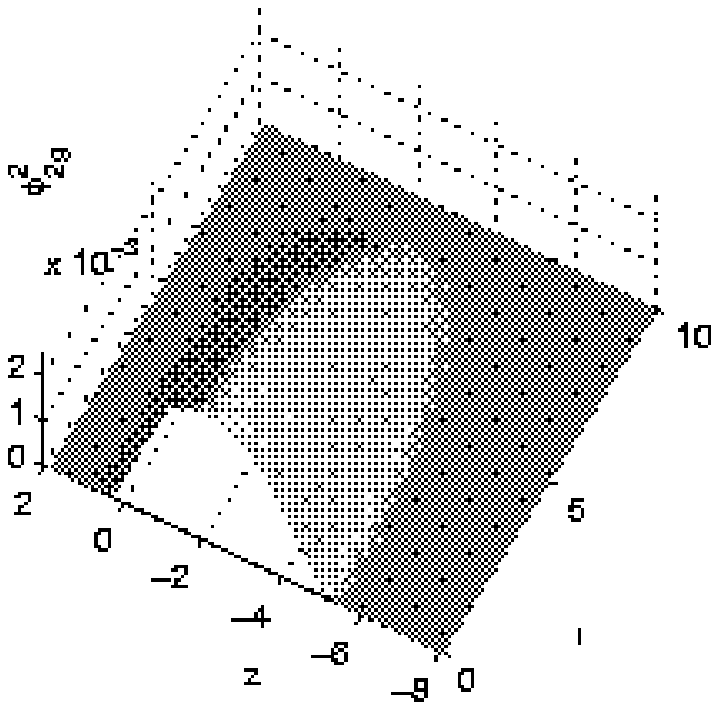,height=6.5cm,width=6.5cm,angle=0}}

Figure 6:  Ground state solution in 3d with cylindrical symmetry 
in example 3 for case II. Condensate wave function on two
lines for $\hat{z}_{0,1}/a_0=-\hat{z}_{0,2}/a_0=0.01, 0.1, 0.5, 2.0, 4.0$. 
On the line $z=0$: a).  
$\phi_{g,1}(r,0)$; c).$\phi_{g,2}(r,0)$. On the line $r=0$:
 b). $\phi_{g,1}(0,z)$; d). $\phi_{g,2}(0,z)$. 
Surface plot of the condensate  density functions for  
$\hat{z}_{0,1}/a_0=-\hat{z}_{0,2}/a_0=2.0$: 
e). $|\phi_{g,1}|^2$; f). $|\phi_{g,2}|^2$.   

\end{figure} 

 \begin{figure}[htb] 
\centerline{a).\psfig{figure=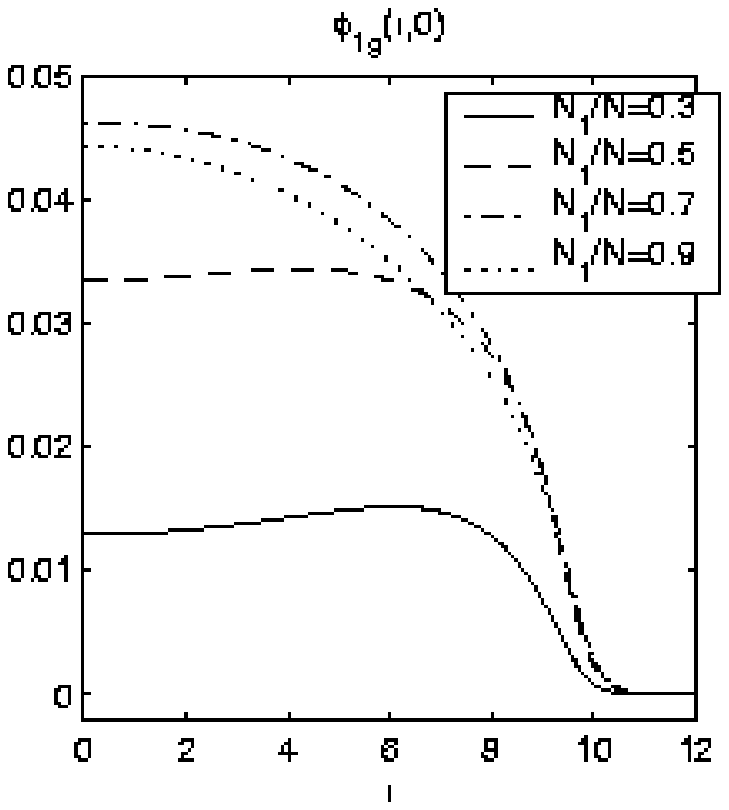,height=6.5cm,width=6.5cm,angle=0}
b).\psfig{figure=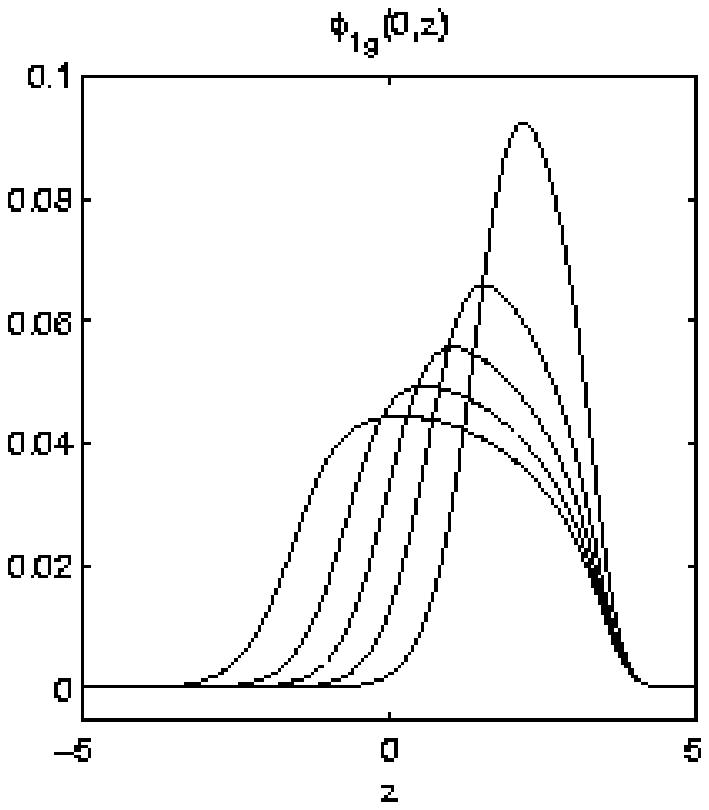,height=6.5cm,width=6.5cm,angle=0}} 
\centerline{c).\psfig{figure=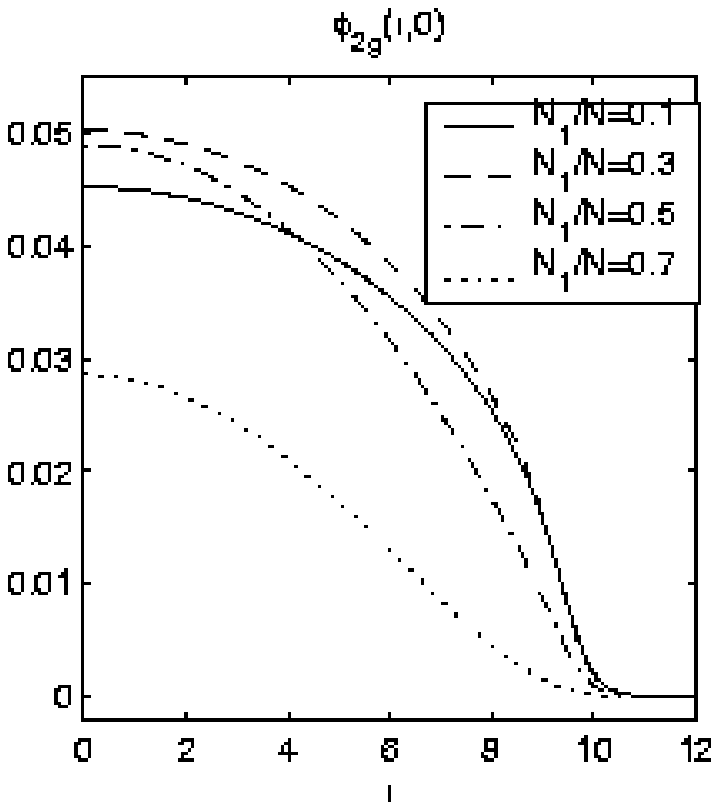,height=6.5cm,width=6.5cm,angle=0}
d).\psfig{figure=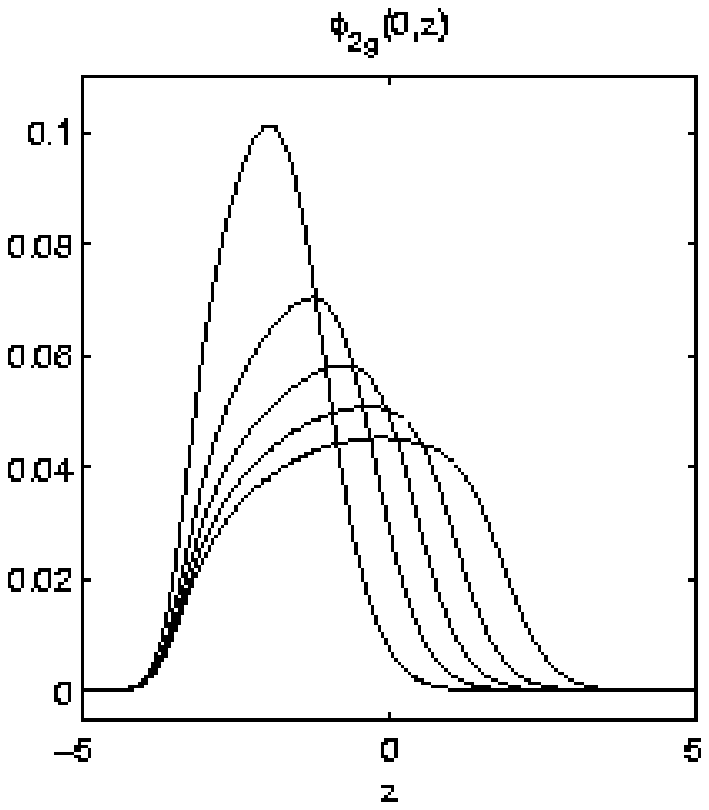,height=6.5cm,width=6.5cm,angle=0}} 
 \centerline{e).\psfig{figure=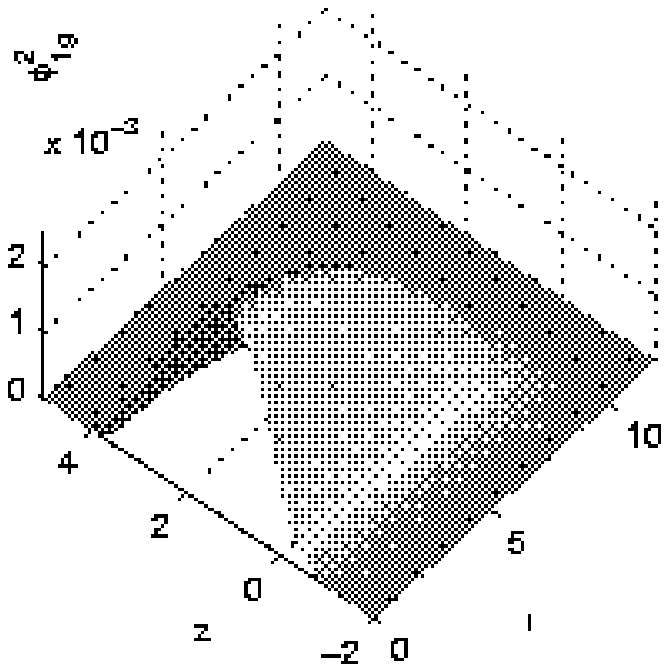,height=6.5cm,width=6.5cm,angle=0}
f).\psfig{figure=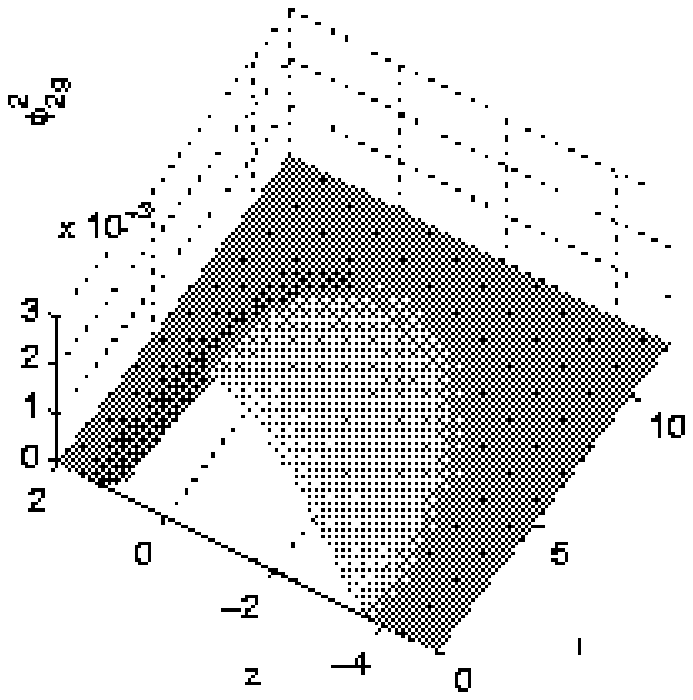,height=6.5cm,width=6.5cm,angle=0}}

Figure 7:  Ground state solution in 3d with cylindrical symmetry 
in example 3 for case III. Condensate wave function on two
lines for $N_1^0/N^0=0.1,0.3,0.5,0.7,0.9$. On the line $z=0$: a).  
$\phi_{g,1}(r,0)$; c).$\phi_{g,2}(r,0)$. On the line $r=0$:
 b). $\phi_{g,1}(0,z)$ (in the order of decreasing peak); 
d). $\phi_{g,2}(0,z)$ (in the order of increasing peak). 
Surface plot of the condensate  density functions for  
$N_1^0=N_2^0=500,000$: 
e). $|\phi_{g,1}|^2$; f). $|\phi_{g,2}|^2$.   

\end{figure}

 \begin{figure}[htb] 
\centerline{a).\psfig{figure=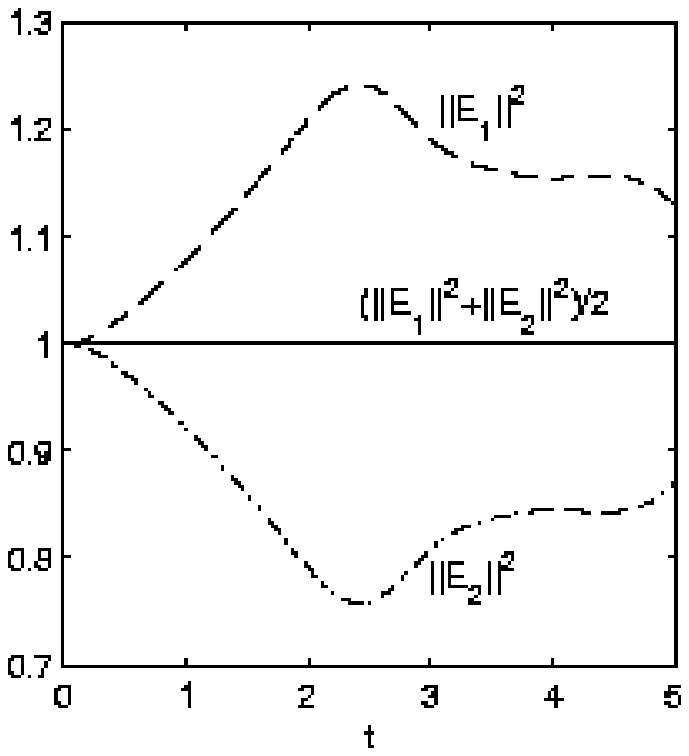,height=6.5cm,width=10cm,angle=0}} 
\centerline{b).\psfig{figure=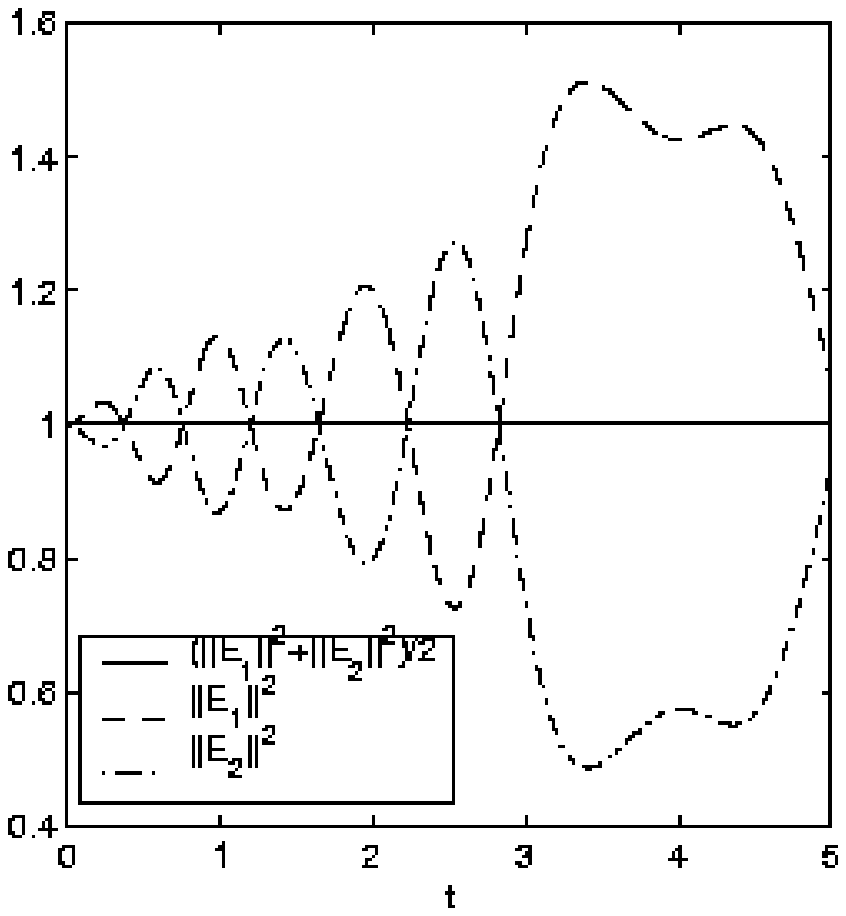,height=6.5cm,width=10cm,angle=0}} 
 \centerline{c).\psfig{figure=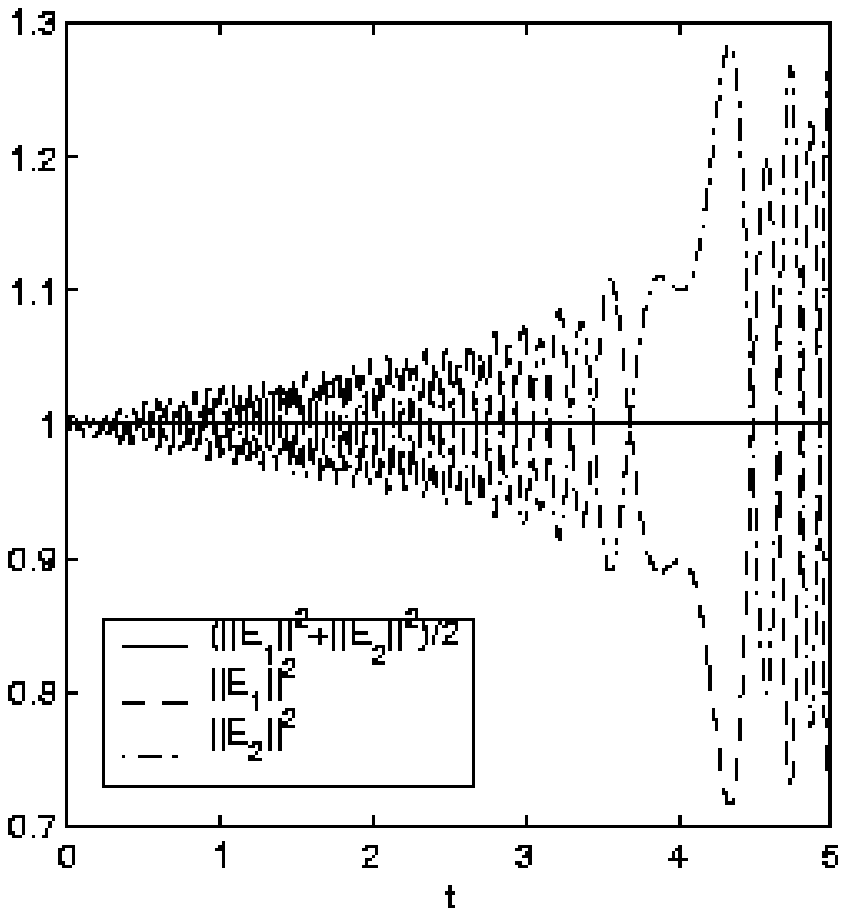,height=6.5cm,width=10cm,angle=0}}

Figure 8:  Time evolution of the mean of the density functions
for the two components, $\|\psi_1\|^2$, $\|\psi_2\|^2$ (labeled as 
$\|E_1\|^2$ and $\|E_2\|^2$, respectively, which 
indicates the time evolution of the number of particles in 
the two components, $N_1^0\|\psi_1\|^2$, $N_1^0\|\psi_2\|^2$, respectively) 
for different driven field frequencies $\Og$. 
a). $\Og=6.5\tm2\pi\;[1/s]$; b). $65\tm2\pi\;[1/s]$;
c). $650\tm2\pi\;[1/s]$.

\end{figure} 

 \begin{figure}[htb] 
\centerline{\psfig{figure=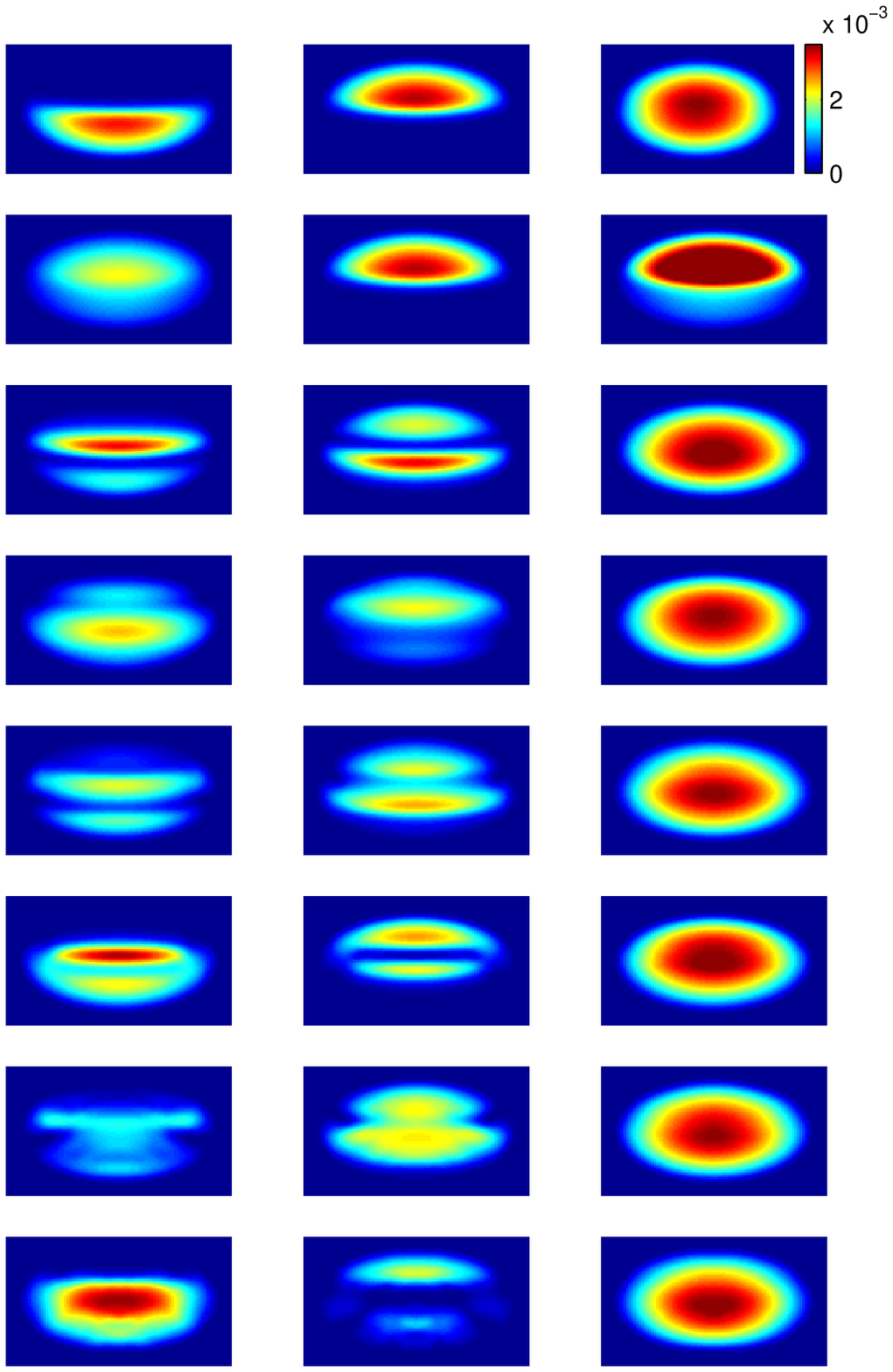,height=19cm,width=15cm,angle=0}} 

Figure 9:  Time evolution of the density functions
for the two components  for  the driven field frequencies 
$\Og=65\tm2\pi\;[1/s]$ at different time, from top to bottom:
$t=0.0,0.24,0.58,0.98,1.42,1.96,2.52,3.4$.
Left column: $|\psi_1|^2$;
middle column: $|\psi_2|^2$, right column: $|\psi_1|^2+|\psi_2|^2$. 

\end{figure} 
\end{document}